\documentclass[
 reprint,
 amsmath,
 pre,
 amssymb,
 nofootinbib,
 aps,
 longbibliography,
]{revtex4-1}

\usepackage[pdftex]{graphicx}
\usepackage{wasysym}
\usepackage{amsfonts}
\usepackage{ifsym}
\usepackage{pifont}
 \usepackage{footmisc}

\makeatletter
\newlength{\apb@width}
\newcommand{\autoparbox}[2][c]{\settowidth{\apb@width}{#2}\parbox[#1]{\apb@width}{#2}}

\makeatother

\newcommand{\namedref}[2]{\hyperref[#2]{#1~\ref*{#2}}}

\renewcommand{\Re}{\mathop{\mathrm{Re}}}
\renewcommand{\Im}{\mathop{\mathrm{Im}}}

\newcommand{\Csphere}{{}^\bullet\kern-1.2pt C}
\newcommand{\Ctorus}{{}^\circ\kern-1.2pt C}

\newcommand{\nn}{\nonumber}

\newcommand{\COMMENT}[1]{}

\newcommand{\neqa}{\nonumber\end{eqnarray}}
\newcommand{\la}[1]{\label{#1}}

\newcommand{\<}{{\langle}}
\renewcommand{\>}{{\rangle}}

\newcommand{\re}{\relax{\rm I\kern-.18em R}}

\newcommand{\gcheck}{\color{ForestGreen}\checkmark}
\newcommand{\rcross}{\color{red}\ding{55}}

\def\su2{{SU(2)}}

\def\[{\left[}
\def\]{\right]}

\def\({\left(}
\def\){\right)}
\def\[{\left[}
\def\]{\right]}

\def\<{\langle}
\def\>{\rangle}

\def\i2{\frac{i}{2}}

\def\2F1{\,_2{\rm F}_1}

\usepackage[usenames,dvipsnames]{xcolor}
\usepackage{color}
\usepackage{graphicx}
\usepackage{dcolumn}
\usepackage{bm}
\usepackage{hyperref}

\usepackage{cancel}
\usepackage{ctable}
\usepackage{booktabs}
\newcolumntype{L}[1]{>{\raggedright\let\newline\\\arraybackslash\hspace{0pt}}m{#1}}
\newcolumntype{C}[1]{>{\centering\let\newline\\\arraybackslash\hspace{0pt}}m{#1}}
\newcolumntype{R}[1]{>{\raggedleft\let\newline\\\arraybackslash\hspace{0pt}}m{#1}}

\newcommand{\beq}{\begin{equation}}
\newcommand{\eeq}{\end{equation}}
\newcommand{\beqq}{\begin{equation*}}
\newcommand{\eeqq}{\end{equation*}}
\newcommand\beqa{\begin{eqnarray}}
\newcommand\eeqa{\end{eqnarray}}
\newcommand\beqaa{\begin{eqnarray*}}
\newcommand\eeqaa{\end{eqnarray*}}
\newcommand\bea{\begin{array}}
\newcommand\eea{\end{array}}

\begin{document}


\title{Where is String Theory?}

\author{Andrea Guerrieri}
\affiliation{School of Physics and Astronomy, Tel Aviv University, Ramat Aviv 69978, Israel}
\affiliation{Instituto de F\'isica Te\'orica, UNESP, ICTP South American Institute for Fundamental Research, Rua Dr Bento Teobaldo Ferraz 271, 01140-070, S\~ao Paulo, Brazil}
\author{Jo\~ao Penedones}
\affiliation{Fields and String Laboratory, Institute of Physics, \'Ecole Polytechnique F\'ed\'erale de Lausanne (EPFL), \\
Rte de la Sorge, BSP 728, CH-1015 Lausanne, Switzerland}
\author{Pedro Vieira}
\affiliation{Perimeter Institute for Theoretical Physics, 31 Caroline St N Waterloo, Ontario N2L 2Y5, Canada}
\affiliation{Instituto de F\'isica Te\'orica, UNESP, ICTP South American Institute for Fundamental Research, Rua Dr Bento Teobaldo Ferraz 271, 01140-070, S\~ao Paulo, Brazil}


\begin{abstract}
\emph{It is in a prime location, stretching all the way to the edge of the garden, separated from the desert by a formidable swamp.} \\

We use the S-matrix bootstrap to carve out the space of unitary, crossing symmetric and supersymmetric graviton scattering amplitudes in ten dimensions. We focus on the leading Wilson coefficient $\alpha$ controlling the leading correction to maximal supergravity. The negative region $\alpha<0$ is excluded by a simple 
 \textit{dual} argument based on linearized unitarity (the desert). A whole semi-infinite region $\alpha \gtrsim 0.14$ is allowed by the \textit{primal} bootstrap (the garden). A finite intermediate region is excluded by non-perturbative unitarity (the swamp). Remarkably, string theory seems to cover all (or at least almost all) the garden from very 
 large positive $\alpha$ -- at weak coupling -- to the swamp boundary -- at strong coupling. 
\end{abstract}

\pacs{Valid PACS appear here}
\maketitle

At large distances gravity is universal. At short distances it is UV completed. 
The first hints of such completions come from the Wilson coefficients (Wcs) governing the low energy effective action. String theory  leads to some values of the Wcs; other theories to other values. As we will illustrate, the S-matrix bootstrap is a powerful quantitative tool to carve out the allowed space of such Wcs and thus learn about the potential UV completions of gravity.\footnote{The application of the S-matrix bootstrap to the study of EFT has been initiated in \cite{FluxTube} and \cite{Pions2} for the simpler study of 2d flux tubes and 4d massless pions respectively.}

To kick off this program we focus on a simpler setup and set out to study the space of $10$ dimensional gravitational theories with maximal supersymmetry. In $d\ge5$ dimensions gravity is IR finite. The main simplification here is however supersymmetry as it allows us to relate scattering of gravitons to the much simpler scattering of its scalar superpartners. 
The two-to-two scattering amplitude of the graviton multiplet for any 10D theory with maximal SUSY takes the form\footnote{See e.g. equation (7.4.57) in \cite{GSW}  and \cite{juan} for a simple general argument. For $\mathcal{N}=(2,0)$ -- as in type IIB superstring theory -- there is a manifestly supersymmetric representation of this prefactor as $\textbf{R}^4= \delta(Q)$ \cite{Boels:2012ie} while for  $\mathcal{N}=(1,1)$ -- as in type IIA -- such SUSY rewriting is not  known as reviewed in \cite{XiYinPaper}. Nonetheless, in both cases (\ref{Adef}) holds. See  \cite{R4PureSpinor} for a covariant representation of this pre-factor in the pure spinor formalism. } 
\beq
\mathbb{A}_{2\to 2}= \textbf{R}^4 A(s,t,u) \,. \la{Adef}
\eeq
By extracting different components of the $\textbf{R}^4$ prefactor sitting in front we get access to the various scattering processes. At low energy $A(s,t,u) \sim 1/stu$ is the 
universal gravity behavior. 
In $\mathcal{N}=(2,0)$ we can scatter the charged axi-dilaton, for instance, by picking an $s^4$ factor from the $\textbf{R}^4$ prefactor thus getting an amplitude  
\beq
T(s,t,u) \equiv s^4 A(s,t,u) = -8 \pi G_N \left(\frac{s^2}{t}+\frac{s^2}{u} \right) + \dots 
\label{Agrav}
\eeq
There are $t$ and $u$ channel poles corresponding to massless graviton exchanges between the charged scalars; there is no $s$ channel pole since these scalars are charged and  thus can not annihilate. The combination $T$ is very important and will be the central object in this paper since unitarity for the super amplitude turns out be equivalent to usual unitarity for this component as explained in appendix~\ref{superU}.\footnote{In $\mathcal{N}=(1,1)$, we have no charged axi-dilaton and $T$  is not an individual component but unitarity for the superamplitude  is nonetheless the same when expressed in terms of $T$ defined through (\ref{Agrav}).}

The Wcs are in the dots in \eqref{Agrav}. More precisely,\footnote{We use    $\ell_P^4= g_s \ell_s^4$ and $8\pi G_N =  \frac{1}{2}(2\pi)^7 g_s^2\ell_s^8= 64 \pi^7 \ell_P^8 $ to facilitate the comparison with String Theory. \label{conventions}}
\beq
\frac{T(s,t,u)}{8\pi G_N=64 \pi^7 \ell_P^8} = s^4\(\frac{1}{stu}+\alpha \,\ell_P^6+ O(s) \) \label{defAlpha}
\eeq
where the $O(s)$  term is a universal one loop contribution, which we work out in appendix \ref{unitAppendix}. At higher orders in $s$ there are
 subleading Wcs and higher loop contributions. 
 Nicely, the first Wilson coefficient $\alpha$ appears at order $s^0$ and can thus be  cleanly separated from the 1-loop contribution. \footnote{This is not always the case.  In pion physics, for example, such separation is more subtle as both effects come in at the same order in the low energy expansion \cite{Pions2}.} 
The coefficient $\alpha$ controls  the (SUSY completion of the) $Riemann^4$ term in the effective action \cite{Gross:1986iv}.
The purpose of this paper is to study the allowed space of $\alpha$ compatible with the S-matrix Bootstrap principles of analyticity, crossing and unitarity of the 2 to 2 scattering amplitude. 

\begin{figure}[t]
\centering
        \includegraphics[scale=0.425]{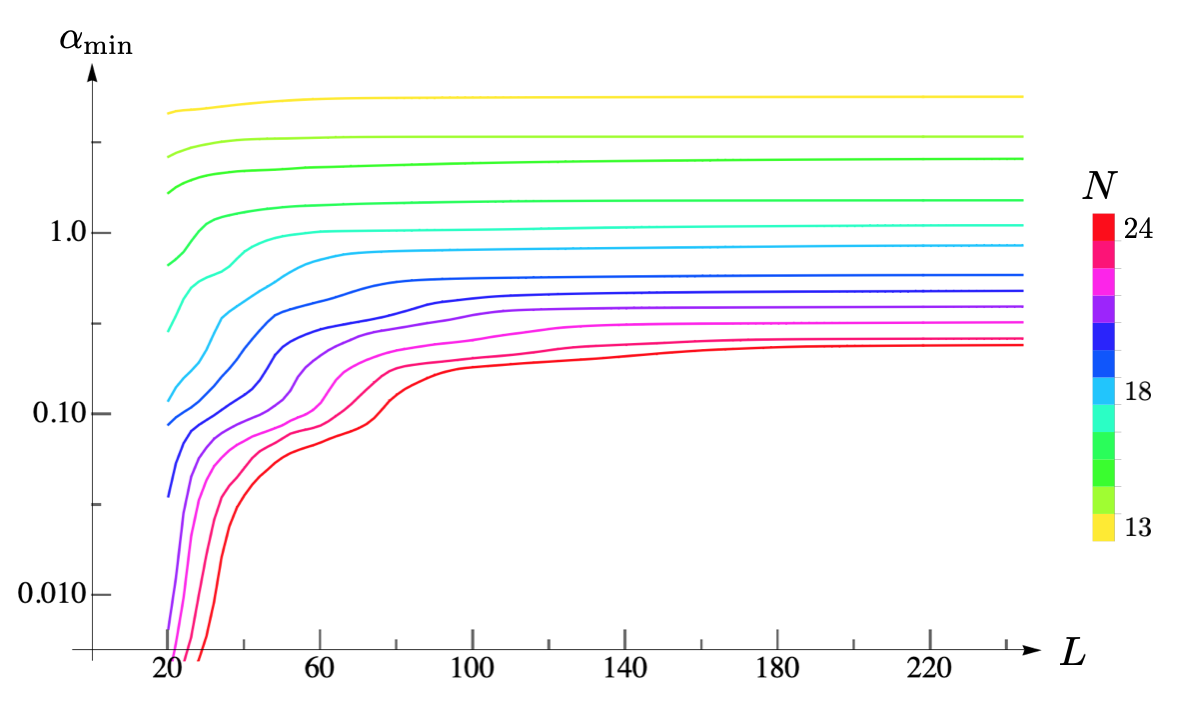}
\caption{Minimum $\alpha_\text{min}(N,L)$. We see that the curves nicely converge towards a plateau whenever $L$ is large enough. The larger $N$ is, the further we need to go in $L$ to reach this plateau. For each value of $N$ we extrapolate these plateaus to estimate $\alpha_\text{min}(N,\infty)$ which we plot in the next figure. 
} 
\label{AllLs}
\end{figure}

In type IIB superstring theory,\footref{conventions} we have \cite{Green:1997tv,Green:1997di,Green:2006gt, Chester:2019jas}
\beq
\alpha^\text{IIB}= \frac{1}{2^6} E_\frac{3}{2}(\tau,\bar{\tau}) \ge \frac{1}{2^6} E_\frac{3}{2}(e^{i \pi/3},e^{-i \pi/3}) \approx 0.1389 
\eeq
where the  non-holomorphic Eisenstein series  depends on the complexified string coupling $\tau= \chi_s +\frac{ i}{g_s}$.  
In fact, it is always larger than a finite positive value (see appendix~\ref{EisensteinAp} for more details).
In type IIA superstring theory, we have (see e.g. \cite{Green:1997di,Green:2006gt,Pioline:2015yea,Binder:2019mpb})
\beq
\alpha^\text{IIA}= \frac{\zeta(3)}{32 g_s^{3/2}} + g_s^{1/2} \frac{\pi^2}{96} \ge \frac{\pi^{3/2} (\zeta(3))^{1/4}}{24\sqrt{3}} \approx 0.1403
\eeq
where the string coupling $g_s \ge 0$.
We conclude that the values realized in String Theory are 
\beq
\alpha  \ge \alpha_\text{min}^\text{ST}\equiv
\frac{1}{2^6} E_\frac{3}{2}(e^{i \pi/3},e^{-i \pi/3}) \approx 0.1389\,. \la{STValues}
\eeq

Our goal is to use the bootstrap to find out the allowed possible values of $\alpha$. How big is the space of allowed quantum gravity UV completions and does string theory fit in this space?

First of all, we can show that $\alpha$ can not be negative. 
Indeed, in appendix  \ref{ArcsAp} we employ the usual contour manipulation arguments \cite{NimaEtAl} -- see also \cite{Arkanihedron,Rattazzipositive,Tolley:2020gtv,simoneft}--  to show that
\beq
 \alpha = \frac{1}{  32 \pi^8 \ell_P^{14}}  \int_{0}^\infty \frac{ds}{s^5}\, {\rm Im}\, T(s+i\epsilon,t=0) \,.
\eeq
The optical theorem then implies
\beq
\alpha \ge 0 \label{dualBound}  \,.
\eeq

%

This is a prototypical example of a rigorous \textit{dual} exclusion bound. No matter how hard we scan over putative ansatze for $T$ -- as we do in the primal formulation -- we will never encounter an amplitude with a negative Wilson coefficient $\alpha$. Beautifully, both type IIA and IIB coefficients reviewed above are indeed always positive.

\begin{figure}[t]
\centering
        \includegraphics[scale=0.39]{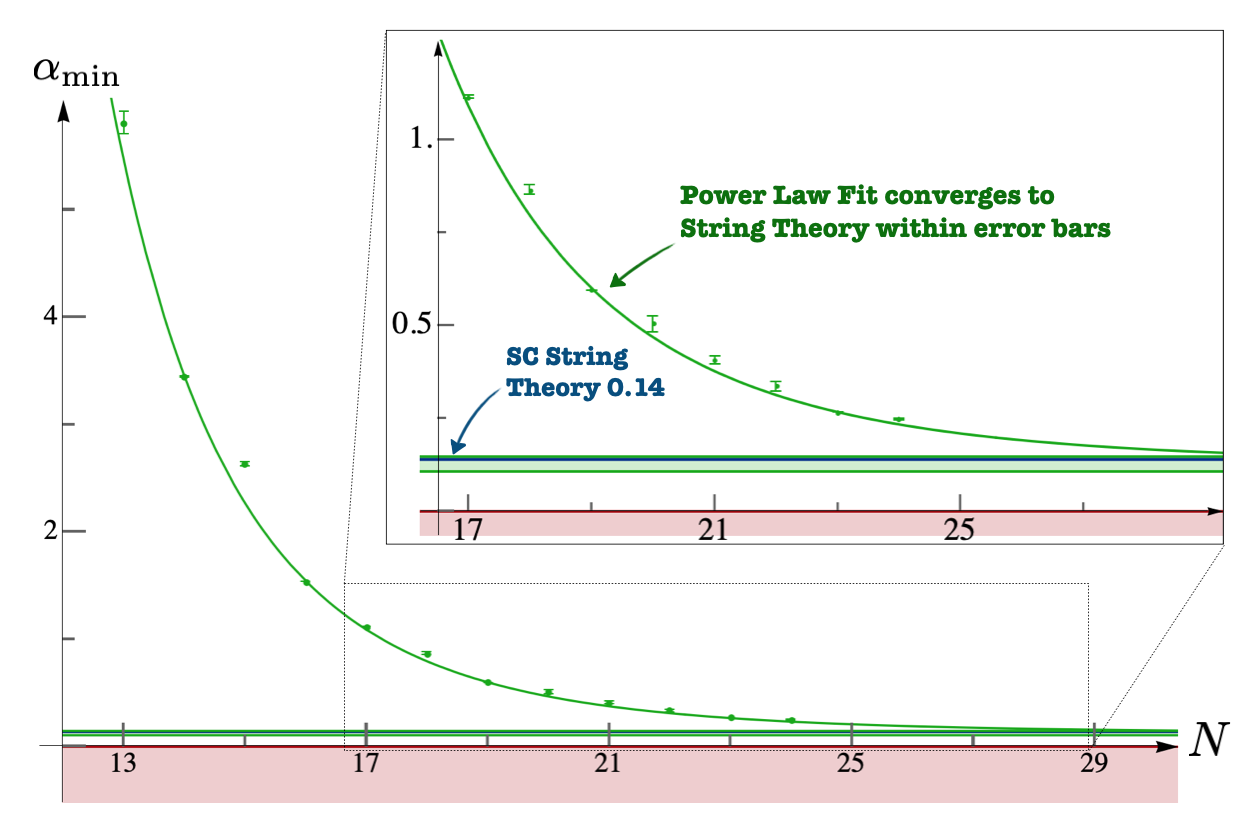}
\caption{Minimum $\alpha_\text{min}(N,\infty)$ obtained by extrapolating the various plateaus in figure \ref{AllNs}. We estimate the error bars here by scanning over a large number of such fits as explained in appendix \ref{numDet}. We then extrapolate these points to estimate $\alpha^\text{Boot}_\text{min}=\alpha_\text{min}(\infty,\infty) \simeq 0.13$ with an uncertainty represented by the green strip. It nicely embraces the strong coupling string prediction depicted by the solid blue line.
} 
\label{AllNs}
\end{figure}

The optimal bound must therefore be somewhere between the dual bound (\ref{dualBound}) and the string theory realization (\ref{STValues}). 
To look for it we turn to the primal S-matrix bootstrap formulation and construct the most general amplitude compatible with maximal SUSY, Lorentz invariance, crossing, analyticity and unitarity following~\cite{4dpaper1,Pions1,Pions2,4dspinning,aninda1,aninda2}. The key representation is given by 
 \beq
\frac{T}{8 \pi G_N}= s^4 \Big( \underbrace{\frac{1}{s t u}}_{\texttt{SUGRA}}+ \underbrace{ \prod_{A=s,t,u}\!\!(\rho_A{+}1)^2\!\!\! \!\!\sum_{a+b+c\le N}^{\prime} \alpha_{(abc)}\rho_s^a \rho_t^b \rho_u^c }_{\texttt{UV completion}}\Big) 
\label{ansatz0}
\eeq
which follows the notation introduced in those references and is discussed in detail in appendix \ref{numDetails} (also \ref{largeSpinAppendix} and \ref{largeEappendix}).

Figure \ref{AllLs} depicts various curves for the minimum value of $\alpha$ for various $N$ (related to the number of parameters in a primal ansatz) as a function of $L$ (maximum spin up to which we impose unitarity of the partial waves) -- see appendix \ref{numDetails} for details. 
We see that as $N$ grows the primal ansatz is capable of minimizing $\alpha$ better and better as expected. But we also see that for each $N$ it is crucial to impose unitarity up to very large spin $L$ to observe convergence of the bound.
For the $N=24$, for instance, we see that we only converge for spin  $L$ around~$220$; for lower $L$  there are important violations of unitarity in partial waves with spin greater than $L$. 
For each $N$ we  extrapolate the curves in figure \ref{AllLs} to estimate the result at~$L=\infty$. Next we fit in $N$ to estimate the~$N=\infty$ final value as depicted in figure \ref{AllNs}. 

\begin{figure}[t]
\centering
        \includegraphics[scale=0.5]{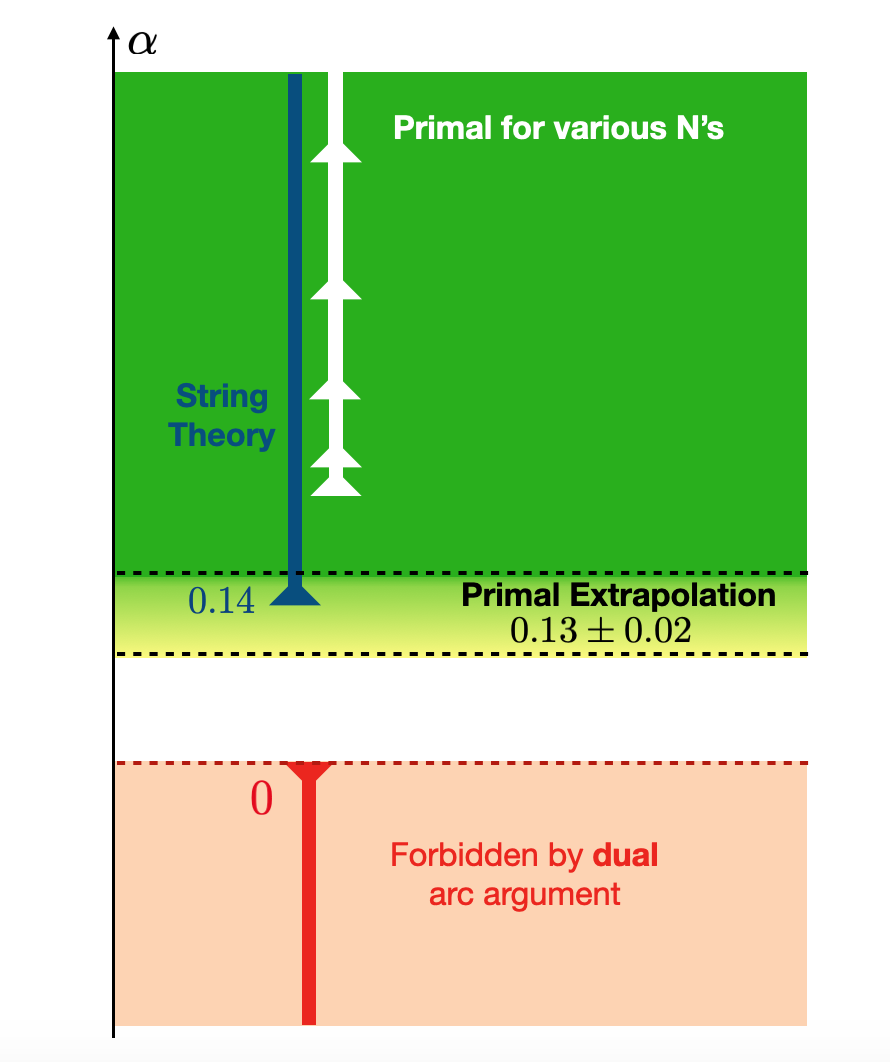}
\caption{String Theory covers all or almost all the allowed quantum gravity theory space.}
\label{SymmaryFigure}
\end{figure}

Note that these fits introduce error bars. More precisely, fitting these curves is a bit of an art as we can a priori pick different number of fitting points and different fit ansatze. We took a large family of plausible fits and weight them by how well they approximate the various numerical points (see appendix \ref{numDet} for details). The spread is an estimate of the final error. In this way we estimate that the $L \to \infty$ extrapolation leads to error bars attached to the points in figure \ref{AllNs} and those error bars, in turn lead to the uncertainty window in the large~$N$ extrapolation denoted by the green shaded region in this figure. In this way, we estimate  that
\beq
\alpha_\text{min}^\text{Boot} \equiv \lim_{N\to \infty \atop L\to \infty } \alpha_\text{min} (N,L)
  \approx 0.13 \pm 0.02 \label{estimate}\,.
\eeq
Comparing with \eqref{STValues} suggests that String Theory realizes all values of $\alpha$ compatible with the S-matrix Bootstrap principles.
It would be useful to increase our numerical precision to check if indeed $\alpha_\text{min}^\text{Boot} = \alpha_\text{min}^\text{ST}$ or if there is some allowed space not realized in Superstring Theory. It would also be fascinating to develop a dual S-matrix bootstrap problem (see e.g. \cite{monolith,2dpaper,talkMartin}) which would extend the simple red excluded region derived above -- the desert -- into the swamp which currently separates it from the green garden included by the primal problem as summarized in  
 figure \ref{SymmaryFigure}. 



\begin{figure}[t]
\centering
        \includegraphics[scale=0.13]{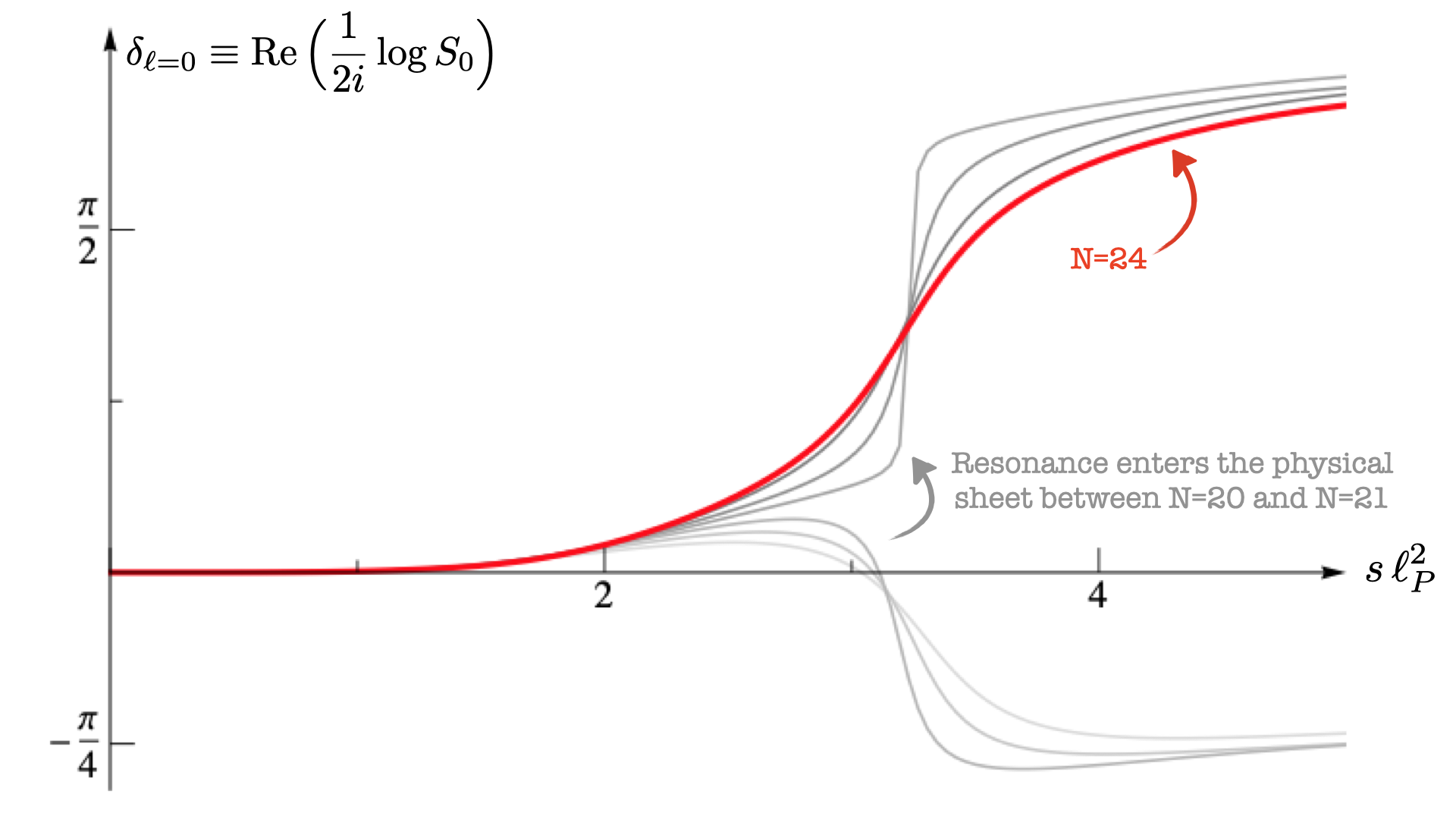}
\caption{Spin zero phase shift as we increase~$N$ from $18,\dots,23$ (in gray) until $24$ (in red). Between $N=20$ and $N=21$ a zero enters the physical sheet -- it is the lightest resonance we encounter. Our best numerics with $N=24$ seem to be close to converging in this energy range and hint at a mass of around $m^2 \ell_P^2 \simeq 3.2 + 0.3 i$.}
\label{spin0resonance}
\end{figure}

As usual with primal problems, it is fascinating to see what physical features the optimal solutions have. In this case, how do phase shifts for theories of quantum gravity living at the boundary of the garden look like? We are investigating this in more detail and hope to report on a more extensive study soon but two fascinating features seem to be robust: (i) There are infinitely many resonances,\footnote{For large spin they seem to lie on a curved Regge trajectory  with~$s_* = m_* ^2 \sim \ell^{3/2}$ as predicted by unitarity, see appendix \ref{largeSpinAppendix}.}
(ii) the lightest resonance is a spin zero resonance which we show in figure \ref{spin0resonance}. 
This scalar resonance is reminiscent of the graviball recently found in~\cite{gravipaper}
using an approximate method to unitarize perturbative amplitudes (see appendix \ref{ap:IAM} for a similar approach in our setup).


We explored here the one dimensional space of $\alpha$, the leading Wilson coefficient. Would be fascinating to explore combined space of  the first two leading Wilson coefficients. What structures do we find in this richer two dimensional space? Is closed superstring theory again in a privileged position? Would be interesting to investigate open strings as well (and possible UV completions of Yang-Mills theory in higher dimensions).

At high energy we expect black holes in any theory of quantum gravity. It would be very nice to see how they fit in our analysis. The Wilson coefficients will likely not be affected dramatically by modifying the high energy inelasticity to the expected behavior but other quantities such as the positions of the various resonances alluded to above might change more significantly.
 
It would also be instructive to see how all this fits in AdS/CFT. 
The analogue of the super-graviton scattering amplitude is the four-point function of the stress tensor multiplet in the dual super conformal field theory.
This has been studied in the large $N$ expansion with maximal supersymmetry  \cite{Aprile,Alday:2017vkk,Aharony:2016dwx,joaoYM, Chester:2020dja,Chester:2020vyz}.
The analogue of our non-perturbative S-matrix Bootstrap is the Superconformal Bootstrap \cite{Beem:2013qxa,  Alday:2013opa, Chester:2014fya, Beem:2015aoa, Beem:2016wfs, Agmon:2017xes}.
It would be interesting to explore this connection in detail. 

It would also be very interesting to repeat our analysis in other dimensions. In 11 dimensions we should make contact with M-theory whose scattering amplitudes have no free parameters. It would also be fascinating to consider less or even no supersymmetry. In that case we have to deal with all the pain and glory of gravitons as spinning particles.\footnote{One will need to generalize the recent work \cite{4dspinning} from 4D to higher dimensions. The formalism developed in \cite{Arkani-Hamed:2017jhn, Chowdhury:2019kaq} should also be useful.}  It would be amazing if the theory of quantum gravity describing our universe would be at a premium location, identifiable through the S-matrix bootstrap.

\begin{acknowledgments}
We thank Francesco Aprile, Nathan Berkovits, Massimo Bianchi, Michael Green, Alexandre Homrich, Shiraz Minwalla, Alessandro Pilloni, Silviu Pufu, Ana Maria Raclariu, Amit Sever and Andrew Tolley for useful discussions. 
Research at the Perimeter Institute is supported in part by the Government of Canada through NSERC
and by the Province of Ontario through MRI. 
This work was additionally supported by a grant from the Simons Foundation (JP: \#488649, PV: \#488661) and FAPESP grant 2016/01343-7 and 2017/03303-1. 
JP is supported by the Swiss
National Science Foundation through the project 200021-169132 and through the National
Centre of Competence in Research SwissMAP. AG
is supported by The Israel Science Foundation (grant number 2289/18).
\end{acknowledgments}


\appendix

\section{Super Unitarity}
\label{superU}
In a SUSY theory we ought to sum over all elements in SUSY multiplets in intermediate states. For example, for intermediate two particle states we have \cite{Bern:1998ug, Green:2008uj, Boels:2012ie}
\beq
\sum_\text{two pt} \textbf{R}^4_{12\to \text{two pt}} \textbf{R}^4_{\text{two pt}\to 34}  = \textbf{R}^4_{12 \to 34} \times s^4 \nonumber
\eeq
so that the unitarity relation for the amplitude $A$ defined in (\ref{Adef}) picks an extra $s^4$ in the right hand side, $$\text{Disc} A  \ge  s^4 \int d\text{LIPS} \,A\times A \,,$$
where $ d\text{LIPS}$ stands for the usual Lorentz invariant phase space integral.
In turn, by multiplying both sides of this relation by $s^4$, we conclude that the axi-dilaton component $T(s,t,u)=s^4 A(s,t,u)$ described in this paper obeys the \textit{usual} ten dimensional unitarity without any additional factors. In terms of partial waves this means that the absolute value of (\ref{projections}) defined below must be smaller than one. 

With $\mathcal{N}=(2,0)$ -- as in type IIB -- we can interpret the necessity of this unitarity condition in another way: the component $T$ describes the scattering of two charged scalars. There are no other two particle states with that much charge so in the intermediate two particle states we can only have those same states flowing. As such unitarity for that component should be just the usual unitarity. Of course, this simple argument shows that the unitarity condition whence derived is necessary but it does not immediately show that it is sufficient. The previous argument is therefore stronger (and besides it holds for both  $\mathcal{N}=(2,0)$ and  $\mathcal{N}=(1,1)$).

We conclude this supersymmetric appendix by recalling that $R^2$ and $R^3$ super symmetrizations do not exist and that is why we jump directly from $1/stu$ to a constant in (\ref{defAlpha}). SUSY tells us we can strip out all polarizations into the prefactor $\textbf{R}^4$  which multiplies a fully crossing symmetric function as  in (\ref{Adef}). At leading order that crossing symmetric function can be computed in SUGRA, has dimension $-3$ in terms of powers of $s$ and is equal to $1/stu$. 

$R^2$ would have dimension $-2$ so it would have to be something like $1/st+1/su+1/tu$ since we want to have at most single poles as singularities. But this combination vanishes since it is proportional to $s+t+u=0$. 

$R^3$ would have dimension $-1$ so it would be of the form $1/s+1/u+1/t$ which no longer vanishes. However, when extracting the axi-dilaton component as in (\ref{defAlpha}) we would generate $s^4/t$ like terms corresponding to massless spin $4$ particles which we do not want. (For $1/stu$ we were producing good spin $2$ graviton exchanges, see \eqref{Agrav}.). Hence we are also forced to kill such terms. 

Altogether this is why we start with $R^4$.

\section{One Loop Unitarization}
\la{unitAppendix}

In this appendix, we start in general spacetime dimension $5\le d \le 11$ with maximal supersymmetry  and determine the one-loop contribution to \eqref{Agrav} using elastic unitarity.

We focus on 2 to 2 scattering of identical charged scalars (in the graviton supermultiplet):  
\beq
T=
N_d \ell_P^{d-2}\frac{s^4}{stu}  \left(1 + \alpha \, \ell_P^6 s t u +
\ell_P^{d-2}f_{1 }(s,t,u) +\cdots \right)
\label{eq:Tscalars}
\eeq
where $N_d$ is a conventional numerical factor (for instance~$N_{10}=64\pi^{7}$) and $f_{1 }(s,t,u)$ represents the 1-loop supergravity contribution.   
As explained in appendix \ref{superU}, unitarity for the full graviton multiplet is equivalent to the usual unitarity condition for this scalar amplitude. The amplitude $T$ has a standard partial wave expansion  
\beq
T= 2 i s^{\frac{4-d}{2}} \sum_{\ell=0 \atop even}^\infty \left(1-e^{ 2i\delta_\ell(s)}\right)
P_\ell^{(d)}\left(\frac{u-t}{u+t}\right)\,,
\label{PWE}
\eeq
with $P_\ell$ proportional to the Gegenbauer polynomials:
\beq
P_\ell^{(d)}(z)=2^{2d-5}\pi^{\frac{d-3}{2}} (d+2\ell-3)\Gamma\left( \frac{d-3}{2} \right) C_\ell^{\frac{d-3}{2}}(z)\,.
\eeq
Equations \eqref{eq:Tscalars} and \eqref{PWE} are compatible if the phase shift has   the following low energy expansion 
\begin{align}
\delta_\ell(s) =  \delta_\ell^{(0)}(\ell_P^2 s)^{\frac{d-2}{2}}  + \dots 
\end{align}
with
\beq
\sum_{\ell=0 \atop even}^\infty  \delta_\ell^{(0)}
P_\ell^{(d)}\left(z\right) = \frac{N_d}{1-z^2}    \,.
\label{treedeltaeq}
\eeq
This gives \cite{Caron-Huot:2018kta} \footnote{\label{footGegen} One can use the integral representation
\beq
C_\ell^{(\lambda )}(z)=\int_0^\pi dx \frac{2^{1-2 \lambda } \Gamma (\ell+2 \lambda
   ) \sin ^{2 \lambda -1}(x) \left(\sqrt{z^2-1} \cos
   (x)+z\right)^\ell}{\ell! \Gamma (\lambda )^2}
\eeq
and commute the sum over $\ell$   with the integral over $x$.}
 \begin{align}
\delta_\ell^{(0)} =  \frac{N_d  \Gamma(d-4)   }{2^{2d-5}\pi^{\frac{d-3}{2}}\Gamma\left( \frac{d-3}{2} \right)  (\ell+1)_{d-4}}  \,.
\end{align}

We can use elastic unitarity to build the amplitude and the phase shift to higher orders. 
Notice that  2 to 3 amplitudes are of order $G_N^\frac{3}{2} \sim \ell_P^{\frac{3}{2}(d-2)}$. 
Therefore, the imaginary part of the phase shift is of order
\beq
{\rm Im}\, \delta_\ell(s) \sim P_{2\to 3} \sim \left( \ell_P^2 s \right)^{\frac{3}{2}(d-2)}\,.
\eeq
Using \eqref{PWE}, we can then compute the imaginary part of $T$ to one loop order,
\beq
{\rm Im}\, T= 4 s^{\frac{4-d}{2}} \!\! \left[ \sum_{\ell=0 \atop even}^\infty \left( \delta_\ell(s)\right)^2
P_\ell^{(d)}\Big(\frac{u-t}{u+t}\Big) + O\!\left( \ell_P^2 s \right)^{\frac{3}{2}(d-2)} \right] \nonumber
\eeq
This leads to 
\beq
{\rm Im}\, f_1 = \frac{4}{N_d} s^{\frac{d}{2}-3} t u   \sum_{\ell=0 \atop even}^\infty \left( \delta_\ell^{(0)}\right)^2
P_\ell^{(d)}\left(\frac{u-t}{u+t}\right)  
\eeq
which can be summed in closed form in a given spacetime dimension.\footref{footGegen}

From now on we focus on $d=10$, where we find 
\begin{align}
\label{Imf1}
{\rm Im}\, f_1 =
\frac{\pi ^3 s^3 }{7680 t^2 u^2}
   &\left[-s t u
   \left(s^2+7 (t-u)^2\right) \right.\\
   &\left.+8 t^5
   \log \left(-t/s\right)+8
   u^5 \log
   \left(-u/s\right)\right]\nonumber
   \end{align}
where we used $N_{10}=64\pi^{7}$.
From this one can use analyticity and crossing to reconstruct the function $f_1$ from its imaginary part,
\begin{align}
&f_1= \frac{\pi ^2 }{960}\left[
   -\frac{s^3 \left(t^5+u^5\right)  \log^2(-s)}{2 t^2u^2}\right. \label{finalf1}
 \\&
   -\frac{t^3 \left(s^5+u^5\right) \log ^2(-t)}{2 s^2 u^2}   -\frac{u^3\left(s^5+t^5\right) \log ^2(-u)}{2s^2 t^2}
     \nonumber \\& \nonumber
   +\frac{s^3 t^3 \log (-s)\log (-t)}{u^2}
   +\frac{s^3 u^3 \log(-s) \log (-u)}{t^2} \\&\nonumber
   +\frac{t^3 u^3\log (-t) \log (-u)}{s^2}
   -\frac{s^4   (s^2+7(t-u)^2 ) \log (-s)}{8 t u}
   \\&\nonumber
   -\frac{t^4 (t^2+7(s-u)^2 ) \log (-t) }{8 s u}
   -\frac{u^4  (u^2+7  (s-t)^2 ) \log ({-}u)}{8 s  t}\\&\nonumber
   \left.
   +\frac{1}{8}
   \left(s^2+t^2+u^2\right)^2
   \Big(1+\frac{\pi ^2 \left(s^6+t^6+u^6-13 s^2 t^2 u^2\right)}{2 s^2
   t^2 u^2}   \Big)
   \right]
\end{align}
One can easily check that this crossing symmetric function has the s-channel imaginary part given by \eqref{Imf1}. The rational part is completely fixed by dimensional analysis and  the requirement that $f_1$ vanishes at $t=0$, so that the residue of the $T$ amplitude at $t=0$ is not affected by the one loop contribution. The reader may worry that we have not included a mass scale to make the argument of the logarithms dimensionless. In fact, this can be done at no cost because the final answer is invariant under the replacement $ \log (-w) \to \log (-w/\mu^2)$ for $w=s,t,u$. This corresponds to the absence of one loop counterterms. We have checked that \eqref{finalf1} agrees with  the 
low energy expansion of the four-particle genus-one amplitude in type~II superstring theory computed in~\cite{Green:2008uj}.

%

\section{Dispersive formula for $\alpha$} \la{ArcsAp}

Consider the function
\beq
g(z) \equiv \left. \frac{T(s,t)}{8\pi G_N s^4}\right|_{s=-t/2+z}
= \frac{1}{t  (\frac{t}{2} +z)(\frac{t}{2}-z)}+ \alpha \,\ell_P^{6} +\dots
\eeq
for fixed $t$.  
This function has branch points at $z=\frac{t}{2}$ and $z=-\frac{t}{2}$. The brach cuts go to infinity along the real axis and $g(z^*)=\left[g(z)\right]^*$ on the first Riemann sheet. Furthermore, crossing symmetry implies that $g(z)=g(-z)$. 
In addition, we assume that $g(z)$ decays when $|z|\to \infty$. Notice that this is weaker than the usual assumption that the   amplitude $T(s,t)/s^2 \to 0$  as $|s| \to \infty$ \cite{Martin:1962rt, Camanho:2014apa, simoneft}.

\begin{figure}[t]
\centering
        \includegraphics[scale=0.46]{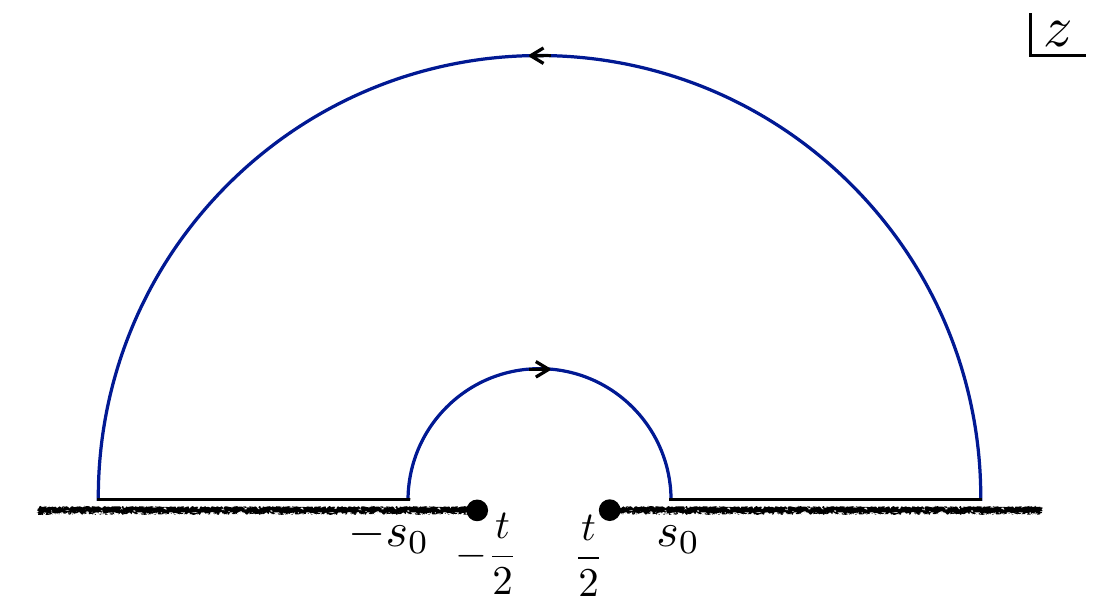}
\caption{ Integration contour leading to equation \eqref{idCauchy}. The large arc does not contribute because $g(z)\to 0$  when $|z|\to \infty$.}
\label{fig:arcs}
\end{figure}

Consider now the contour integral in figure \ref{fig:arcs} for the function $g(z)/z$.
Cauchy theorem implies that
\begin{align}
i \int_0^\pi d\theta\, g ( s_0 e^{i\theta}  )  &=\int_{s_0}^\infty \frac{dz}{z} \left[ g(z+i\epsilon) - g(-z+i\epsilon) \right] \nonumber \\
&=2 i \int_{s_0}^\infty \frac{dz}{z} {\rm Im}\, g(z+i\epsilon)  
\label{idCauchy}
\end{align}
We now consider this equation in the limit 
\beq
0<|t|<s_0\to 0 \,.
\eeq
Consider first the left hand side of \eqref{idCauchy},
\beq
i   \int_0^\pi d\theta \left[
\frac{1}{t  (\frac{t}{2} +z)(\frac{t}{2}-z)} +\alpha\, \ell_P^6 +\dots \right]_{z= s_0 e^{i\theta}}\nonumber
\to i \pi   \alpha \, \ell_P^6
\eeq
where we used that the next  terms in the low energy expansion vanish when $0<|t|<s_0\to 0$. 
In the same limit, the right hand side of \eqref{idCauchy} gives
\beq
 \frac{2 i }{  8 \pi G_N}  \int_{0}^\infty \frac{ds}{s^5}\, {\rm Im}\, T(s+i\epsilon,t=0) \,.
\eeq
Therefore, we conclude that
\beq
 \alpha =  \int_{0}^\infty d(\ell_P^2 s)\, \frac{ {\rm Im}\, T(s+i\epsilon,t=0) }{32 \pi^8 \ell_P^{16} s^5  } \,.
\label{alphasumrule}
\eeq
Here we used $8\pi G_N=64 \pi^7 \ell_P^8$.
Notice that the low energy behavior ${\rm Im}\, T(s+i\epsilon,t=0) \sim s^{5}$  ensures IR finiteness of the integral.

\begin{figure}[t]
\centering
        \includegraphics[scale=0.176]{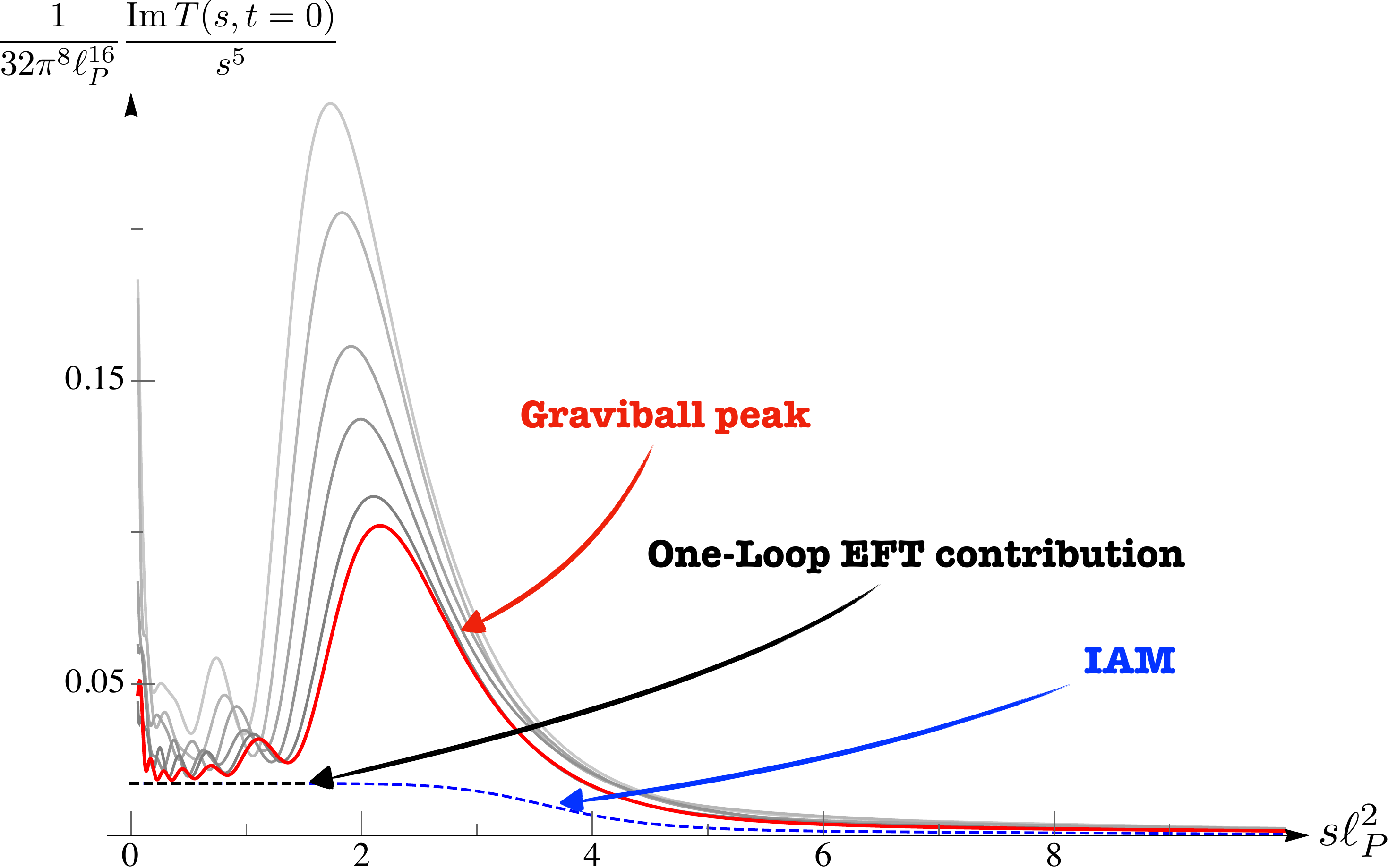}
\caption{Integrand of the sum rule for $\alpha$ in \ref{alphasumrule} as we increase $N$ from 19 to 23 (in gray) until $N=24$ in red. The dashed black line is the value of the integrand from the one-loop EFT prediction $\pi^2/576$. The dashed blue line is the IAM approximation discussed in appendix \ref{ap:IAM}. The value of $\alpha$ is dominated by the spin-0 graviball region where the momentum expansion breaks down.} 
\label{ImTintegral}
\end{figure}

It is quite instructive to plot this integrand to see which energy range is contributing most to $\alpha$.\footnote{We thank A.Tolley for suggesting this analysis and for illuminating exchanges.} 
In figure \ref{ImTintegral} we plot it for the last six $N$'s in our numerics. We see that the dominant contribution to the value of $\alpha$ comes from the strong coupling region around the graviball mass. Close to 
threshold, however, the integrand seems to converge to the EFT one-loop prediction as we increase $N$. 

\section{Superstring theory} \la{EisensteinAp}

\begin{figure}[t]
\centering
        \includegraphics[scale=0.34]{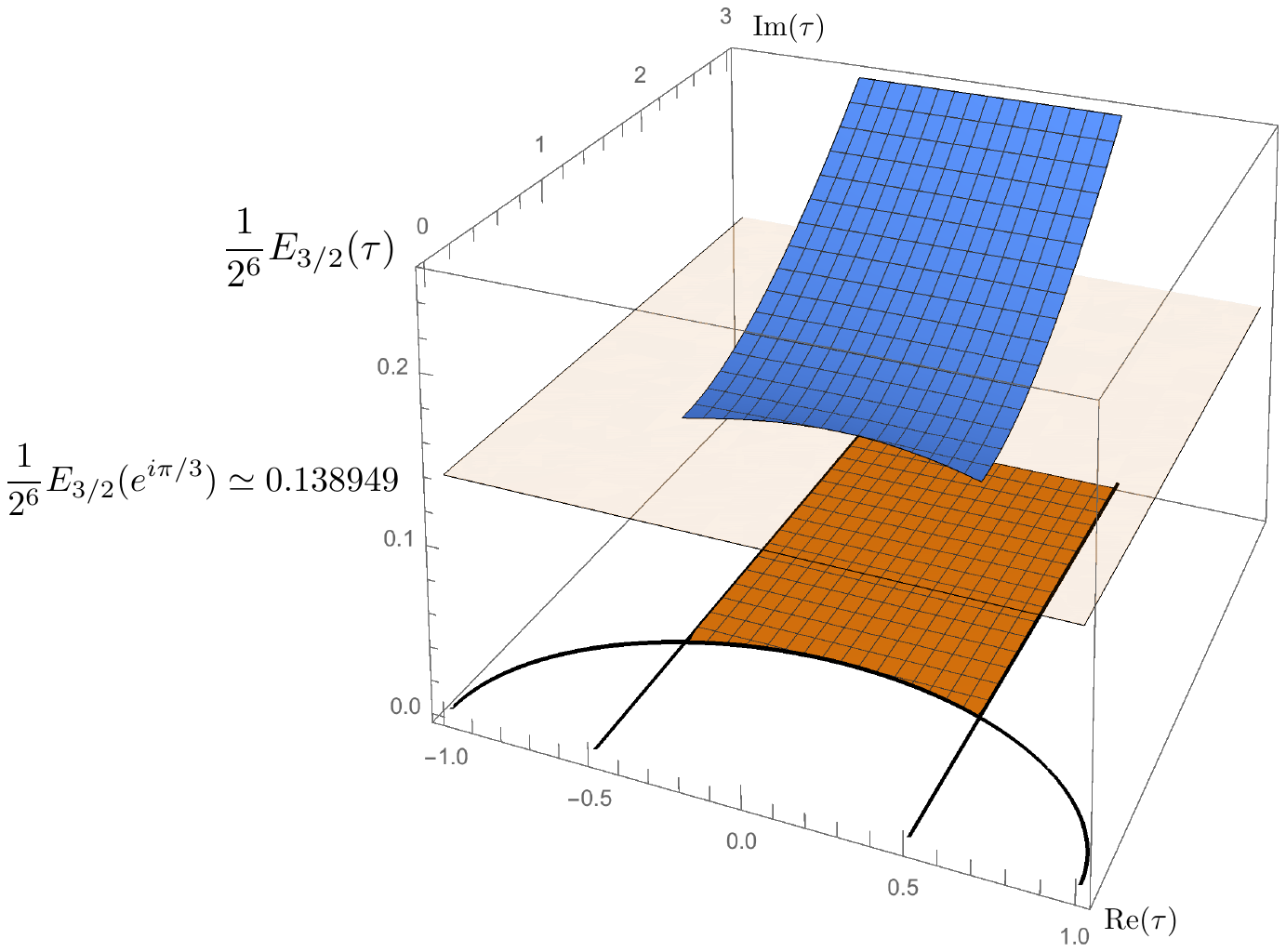}
\caption{In type IIB superstring theory, the leading Wilson coefficient is given by the non-holomorphic Eisenstein series: $\alpha^{\text{IIB}} = \frac{1}{2^6} E_{\frac{3}{2}}(\tau,\bar{\tau})$ . It takes its lowest values at the corners $\tau=e^{i \pi/3},e^{2i \pi/3}$ of the fundamental domain. 
The minimum value at these corners is approximately equal to $0.14$. The coefficient $\alpha^{\text{IIB}}$ as predicted from string theory can thus span any value larger than $0.14$; at weak coupling $\tau \to i\infty$ and $\alpha$ takes very large values.
 } 
\label{Eisenstein}
\end{figure}

In type IIB superstring, we have \cite{Green:1997tv, Chester:2019jas}
\beq
\alpha^\text{IIB}= \frac{1}{2^6} E_\frac{3}{2}(\tau,\bar{\tau})  
\eeq
where the  non-holomorphic Eisenstein series reads
\beq
E_\frac{3}{2}(\tau,\bar{\tau})  = \sum_{m,n \in \mathbb{Z} \atop (m,n)\neq (0,0)} \frac{\left({\rm Im}\, \tau\right)^\frac{3}{2}}{|m\tau+n|^3}\,.
\eeq
This function is invariant under  $SL(2,\mathbb{Z})$ transformations
\beq
\tau \to \frac{a\tau +b}{c\tau +d}
\eeq
with $a,b,c,d \in \mathbb{Z}$ and $a d -b c =1$.
Therefore, it is sufficient to plot it over the fundamental domain 
$\left| {\rm Re}\, \tau \right| \le 1$ and $|\tau|\ge 1$, as shown in figure \ref{Eisenstein}.
The minimal value 
is attained at the "corners" $\tau=e^{i\pi/3}$ and  $\tau=e^{2i\pi/3}$.
This value can be written in terms of Hurwitz zeta function~\cite{AldayBissi} 
\beq
E_\frac{3}{2}(e^{i\pi/3},e^{-i\pi/3})=
\frac{ 3^{\frac{1}{4}} \zeta
    (\frac{3}{2 })
   \left[\zeta
   \left(\frac{3}{2},\frac{1}{3}\right
   )-\zeta
   \left(\frac{3}{2},\frac{2}{3}\right
   )\right]}{ \sqrt{2}}\,.
\eeq

\begin{figure}[t]
\centering
        \includegraphics[scale=0.455]{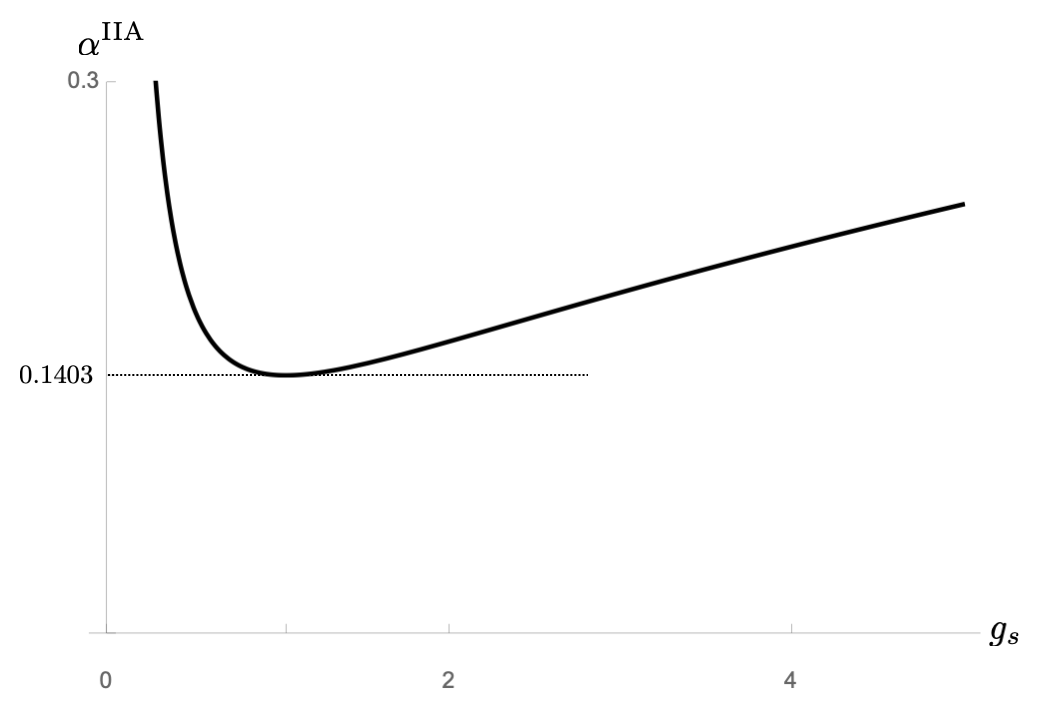}
\caption{The leading Wilson coefficient $\alpha^\text{IIA}$ as a function of the string coupling in type IIA superstring theory. All values above the minimum are allowed and $\alpha^{\text{IIA}}$ diverges at weak coupling.
 } 
\label{alphaIIA}
\end{figure}

In type IIA superstring theory, we have (see e.g. \cite{Binder:2019mpb})
\beq
\alpha^\text{IIA}= \frac{\zeta(3)}{32 g_s^{3/2}} + g_s^{1/2} \frac{\pi^2}{96}  
\eeq
In figure \ref{alphaIIA}, one can see that this function attains its minimum value 
$\frac{\pi^{3/2} (\zeta(3))^{1/4}}{24\sqrt{3}}\approx 0.14$
for 
$g_s^2 = 9\zeta(3)/\pi^2$.

Notice that the perturbative expansions are the same in type IIA and IIB.
Indeed, expanding the Eisenstein series at weak coupling $g_s=1/{\rm Im}\, \tau$ one finds (see e.g.   appendix A of \cite{Chester:2019jas})
\beq
\alpha^\text{IIB}= \alpha^\text{IIA} +O
\left(e^{-\frac{2\pi}{g_s}}\right)  \,.
\eeq
This partially explains  the coincidence $\alpha^\text{IIA}_\text{min} \approx \alpha^\text{IIB}_\text{min} $.

\section{Ansatz Details} \la{numDetails}
Following \cite{4dpaper1,Pions1,Pions2}, we write a general ansatz for the amplitude $T$ in (\ref{Agrav}) as a sum over the tree level contribution plus a generic UV completion as (in the rhs we are setting $\ell_P=1$, that is we measure all Mandelstam variables in units of the Planck length)
 \beq
\frac{T}{8 \pi G_N}= s^4 \Big( \frac{1}{s t u}+ \prod_{A=s,t,u}(\rho_A{+}1)^2 \sum_{abc}^{\prime} \alpha_{(abc)}\rho_s^a \rho_t^b \rho_u^c \Big) 
\label{ansatz}
\eeq
where 
\beq
\rho_s \equiv \frac{\sqrt{s_0}-\sqrt{- s}}{\sqrt{s_0}+\sqrt{- s}}
\eeq
maps each of the Mandelstam cut planes to a unit disk.\footnote{The parameter $s_0$ specifies the center of the Taylor expansion in \ref{ansatz}. In our numerics we choose $s_0 \simeq 0.7$.} The prime in the sum stands for the elimination of several terms in this sum: First of all, because of the on-shell condition $s+t+u=0$ the various monomials are not independent \cite{4dpaper1}. We can use this gauge symmetry to kill basically one monomial per total sum $a+b+c$. 

%
%
%
The prefactor
\beq
(\rho_s{+}1)^2(\rho_t{+}1)^2(\rho_u{+}1)^2 \sim \frac{1}{s t u}+ o(1/stu) \label{prefactor}
\eeq
ensures the behavior at infinity is under control. We further eliminate 28 constants $\alpha_{(abc)}$ by requiring the large energy behavior of the partial waves to be compatible with unitarity, as explained in
appendix \ref{largeEappendix}. In (\ref{ansatz}) we sum over all $a+b+c \le N $. Once we take care of the on-shell gauge redundancy and large energy behavior we are still left with quite a lot of free constants ($64$ for $N =13$ all the way to $273$ for $N =24$ say).

Let us stress that the ansatz above respect the expected crossing (it is given by $s^4$ times a fully symmetric function), has the right $t=0$ singularity (coming from the tree level pole exchange alone) and has a controlled large energy behaviour. 

Note that a priori we could relax the high energy conditions; after all, the rho variables can easily generate decaying terms at infinity as explained in \cite{Pions2} so the ansatz could dynamically accommodate for any large energy behaviour. In practice, however, since we impose unitarity at a fixed grid it is better to treat infinity separately.  

Given the ansatz (\ref{ansatz}) we compute the corresponding partial waves
\beq
S_\ell(s)=1+\frac{i}{3 \cdot 2^{18} \pi^4}s^3\int_{-1}^1 dx (1{-}x^2)^3 \frac{C_\ell^{7/2}(x)}{C_\ell^{7/2}(1)}T(s,x) \,, \label{projections}
\eeq
where $t,u=-s/2(1 \mp x)$, 
and impose unitarity -- that is~$|S_\ell(s)|\le 1$ -- for $\ell=0,2,4,\dots, L $ and for a large number of $s>0$ points. More precisely, using the mapping of $s$ into the unit disk $\rho_s$ we chose a set of $n_\text{pts}= 1000$ points at the boundary of this unit disk as 
\beq
\rho(s_k) = e^{i \frac{\pi}{2}(1+\cos \frac{\pi k}{n_\text{pts}+1})} \, ,\quad k=1,\dots, n_\text{pts}
\eeq
which go all the way from small energies $\rho \sim 1$ to high energy $\rho \sim -1$. 
We then took a subset of these points (with around $300$ points per spin) and run our numerics in that grid. We then tried another subset and found basically identical conclusions. This convinces  that the grids we used are thin enough. 

We also impose $\Im T(s,t=0)\geq 0$ which of course follows automatically from unitarity if we impose it for all partial waves, but since we are truncating in spin this is a powerful extra constraint which helps convergence quite a bit.\footnote{In v1 of this paper we did not impose this condition. The curious reader can open that version and note how all plots are nicer in this v2 and how all error bars are also shrunk. We are currently working on pushing N further to confirm the new improved error bars.}

Of course, we still need to extrapolate to large $L $ and~$N$ and that can be challenging. This is discussed in appendix \ref{numDet}.

\section{Large Spin Conditions} \la{largeSpinAppendix}

At large spin, the graviton exchange term in the ansatz behaves as 
\beq
S_\ell(s)=1+i\frac{s^4 \pi^3}{4 \ell^6}+\mathcal{O}(1/\ell^7).
\eeq
This term will generate $\mathcal{O}(1)$ unitarity violations at energies around $s \sim \ell^{3/2}$.
It is clear that all spin modes of our numerical ansatz will need to contribute at sufficiently higher energy in order to recover unitarity.

We find useful to impose a certain amount of unitarity when the spin is asymptotically large.
One way to do it is to impose the simple necessary condition $\Re S_\ell(s) \leq 1$ 
for~$\ell \to \infty$.
This condition is linear in the free parameters of the ansatz and it has a very simple form which was derived in \cite{4dpaper1} for a gapped theory.
When projecting our ansatz (\ref{ansatz}) into partial waves in (\ref{projections}) we end up computing integrals of the form 
\beqa
I_\ell^{bc}(s)=\int\limits_{-1}^1 dx\, (1{-}x^2)^3 \frac{C_\ell^{7/2}(x)}{C_\ell^{7/2}(1)}\rho_t^b\rho_u^c 
\eeqa
where $t,u=-s/2(1\mp x)$. At large $\ell$, 
\beq
I_\ell^{bc}(s) \simeq \frac{1}{\ell^{9}} \times 5040 \sqrt{\frac{s}{s_0}}\, (b  \rho_{-s}^c +c \rho_{-s}^b ) \,.
\label{largespin}
\eeq
Note that very nicely the $\ell$ dependence stands out independent of $b$ and $c$. As such we obtain a single spin independent condition 
\beq
1-\Re S_\ell(s) \propto  \sum^\prime_{abc} \alpha_{(abc)} (b+1)\Im \rho_s^a\rho_{-s}^c \geq 0\,, \label{linear}
\eeq
which is indeed basically the condition from \cite{4dpaper1} up to some simple shifts in the $a,b,c$ indices coming from the prefactor (\ref{prefactor}). To tame (extremely) large spin we impose the linear condition (\ref{linear}) for all points in our grid (or more).

\section{Large Energy Conditions}
\label{largeEappendix}
Boundedness of partial waves for $s\to\infty$ puts strong constraints on the high energy behavior of our numerical ansatz \ref{ansatz}. The $s^4$ factor sitting in front in (\ref{ansatz}) is enhanced by the extra ten dimensional phase space factor $s^3$ in the projections (\ref{projections}) so that in total we get an overall $s^7$ growth in the partial waves which needs to be cancelled since the partial waves' modulus is smaller than $1$ by unitarity. This leads to quite a large number of conditions. Indeed, our projections will be made of integrals of the form \footnote{In this appendix we choose $s_0=1$ in the definition of the $\rho$ variables to make the discussion simpler without loosing generality.}
\beq
I_\ell^{abc}(s)=\rho^a(s)\int_{-1}^1 \mu_\ell^{(10)}(x) \rho(t(s,x))^b\rho(u(s,x))^c\,dx,
\eeq
where
\beq
\mu^{(10)}_\ell(x)=(1-x^2)^3 6!\frac{C_\ell^{(7/2)}(x)}{(\ell+1)_6}.
\eeq
These integrals behave as 
\beq
I_\ell^{abc}=\sum_{i=0}^{14}g^\ell_i(a,b,c) \frac{1}{s^{i/2}}+\sum_{i=8}^{14}h_i^\ell(a,b,c)\frac{\log(s)}{s^{i/2}}+\mathcal{O}\big(\frac{1}{s^{15/2}}\big) \label{goalExpansion}
\eeq
imposing unitarity at infinity energy will thus require the cancellation of (linear combinations of) all the log terms $h$ plus all the powers $g$ up to thirteen (it turns out that imposing these conditions to all spins boils down to imposing $28$ independent conditions).
In addition, we also impose the boundedness of the constant at infinity in $S_\ell(\infty)$, related to $g^\ell_{14}$, for every even spin $\ell \le L$.
In this appendix we explain how to efficiently expand the integrals in the UV to compute all these constants $g$ and $h$ which will enter these large energy conditions.

%
Since $\ell$ is  even, we can use the $x\to -x$ symmetry of the integral and integrate in a smaller region
\beq
I_\ell^{abc}(s)=\rho^a(s)\int_0^1 (\rho(t)^b\rho(u)^c+\rho(t)^c\rho(u)^b)\mu_\ell^{(10)}(x)\,dx.\nonumber
\eeq
Since the integration measure is a polynomial of degree $6+\ell$ that we can expand in powers of $1-x$
\beq
\mu_\ell^{(10)}(x)=\sum_{n=3}^{6+\ell} \mu_n^\ell (1-x)^n\,, \label{measureDecomp}
\eeq
we can reduce the computation of the integrals $I$ at large energy to that of the integrals 
\beq
J^{bc}_n(s)=\int_0^1 \rho(t)^b\rho^c(u)(1-x)^n\,dx \,,
\eeq
which we now turn to. 

\begin{table}
\begin{ruledtabular}
\begin{tabular}{ccccccccc}
 &$ s^{-4}$ & $s^{-9/2}$ &$s^{-5}$
 &$s^{-11/2}$  & $s^{-6}$ &$s^{-13/2}$ & $s^{-7}$\\
\hline
$n=3$& \rcross & \rcross & \rcross
& \rcross & \rcross & \rcross & \rcross \\
$n=4$& \gcheck & \gcheck & \rcross & \rcross & \rcross &  \rcross & \rcross \\
$n=5$& \gcheck & \gcheck & \gcheck & \gcheck & \rcross & \rcross & \rcross \\
$n=6$& \gcheck & \gcheck & \gcheck & \gcheck & \gcheck & \gcheck & \rcross 
\end{tabular}
\end{ruledtabular}
\caption{\label{tab:table1} For $3\leq n\leq 6$ the expansion of the integrand in $J_n^{bc}(s)$ generates divergences. 
There are 16 diverging contributions that we denote with a red crosscheck. 
For $n\geq 7$ the expansion up to order $\mathcal{O}(s^{-7})$ is always finite. 
The asymptotic expansion of the generic $J_{bc}^n(s)$ is therefore trivial provided we analyze separately these lower orders.
 }
\end{table}

The large $s$ expansion of the integrand and the integration do not commute since the large $s$ expansion is effectively an expansion in $1/\sqrt{s(1-x)}$ and $1/\sqrt{s(1+x)}$. We thus need to proceed with care. 
Since we need to go up to order $\mathcal{O}(s^{-7})$ there is only a finite number of dangerous terms. For instance, for $n\geq 7$, the expansion of the integrals can be integrated 
without generating divergences up to the order we are interested in.
Moreover, since $n\geq 3$, the first divergent term will appear at order~$\mathcal{O}(s^{-4})$.
In Table \ref{tab:table1} we list all the integrals that need to be computed separately by the procedure explained in the next section.

\subsection{A simple example of matching}
Let us illustrate how to evaluate a the large $s$ expansion of our integrals on a simple example 
$$j(s)=\int_{0}^1 dx \,\rho(t(s,x))\rho(u(s,x)) \,.$$
%
For large $s$ the integrand has the expansion
\beq
\rho_t \rho_u\simeq-1+\frac{2\sqrt{2}(\sqrt{1{-}x}+\sqrt{1{+}x})}{\sqrt{s(1-x^2)}}+\frac{8(1+\sqrt{1{-}x^2})}{s(1-x^2)}).\nonumber
\eeq
Here we notice that already at the order $1/s$ the integration of the naive expansion of the integrand will fail in estimating the asymptotic $s$ behavior. 

To proceed we thus split the integral in two regions
\beq
j(s)=\underbrace{\int_{0}^{1-\delta} \rho_t \rho_u dx}_{\texttt{outer}} + \underbrace{\int_{1-\delta}^1 \rho_t \rho_u dx}_{\texttt{inner}}
\eeq
where $\delta \ll 1$ is an arbitrary cut-off which must drop in the end. We denote the first/second term as the  {outer}/ {inner region}. 

For the outer region we can just expand the integrand at large $s$ and integrate
\beqa
\texttt{outer} &\simeq& \,1-\delta +4\sqrt{\frac{2}{s}}(\sqrt{\delta}-\sqrt{2-\delta})\nonumber\\
&&+\frac{8}{s}(\arcsin(1-\delta)+\text{atanh}(1-\delta)) \label{outer} \,.
\eeqa
In the inner region we zoom in close to the end point by changing variables $x=1-2\epsilon^2/s$ so that 
\beqa
\texttt{inner} \simeq \frac{4}{s}\int_0^{\Delta}\frac{1-\epsilon}{1+\epsilon}\epsilon \,d\epsilon\nonumber=\frac{2}{s}\left[\Delta (\Delta-4)- 4\log(1+\Delta)\right],
\eeqa
with $\Delta=\sqrt{\delta s/2}$.
Expanding the resulting integral for large $\Delta$ yields 
\beq
\texttt{inner} \simeq -\delta+4\sqrt{\frac{2\delta}{s}}-\frac{4}{s}\log{\frac{\delta\,s}{2}}. \label{inner} 
\eeq
Adding up (\ref{outer}) and (\ref{inner}) we see that the cut-off dependence nicely drops out as it ought to up to order $\mathcal{O}(s^{-1})$ and we get a final result
\beq
j(s) \simeq 1-\frac{8}{\sqrt{s}}+\frac{4}{s}\log{s}+\frac{4 \pi}{s}.\\
\eeq
The appearance of a logarithmic correction at the order where the integral of the expansion diverges is a prototypical example of the more general expansion (\ref{goalExpansion}) as discussed below.

%
%
\subsection{The expansion of $J^{bc}_n(s)$ integrals}

Let us consider  $J^{bc}_n(s)$ with $3\leq n \leq 6$.
We will proceed as done in the previous example: we split the integration domain into an outer and an inner region
\beq
J^{bc}_n(s)=\left(\int_{0}^{1-\delta}+\int_{1-\delta}^1 \right) \rho_{t(s,x)}^b\rho_{u(s,x)}^c (1-x)^n dx.\nonumber
\eeq

In the outer region we can just expand the integrand and compute the integral of the expansion. 

In the inner region we change variables to~$x=1-2\epsilon^2/s$
\beqa
\int\limits_{1-\delta}^1 \rho_t^b\rho_u^c (1-x)^n dx=\frac{2^{n+2}}{s^{n+1}}\int\limits_0^\Delta (\rho_{-\epsilon^2})^b (\rho_{-s+\epsilon^2})^c \epsilon^{2n+1}d \epsilon. \nn
\eeqa
Since $n\geq 3$ this integral will contribute at order $\mathcal{O}(s^{-4})$.
We can further expand the term $(\rho_{-s+\epsilon^2})^c$ for large $s$ up to order $s^{-3}$. This expansion is polynomial in $\epsilon^2$ so that, in the end, we just need to compute (the large $\Delta$ expansion of) integrals of the form
\beqa
\mathcal{J}^{b}_\alpha=&\int_0^\Delta \rho(-\epsilon^2)^b \epsilon^\alpha d\epsilon=\int_0^\Delta \left(\frac{1-\epsilon}{1+\epsilon}\right)^b \epsilon^\alpha d\epsilon \,. \nonumber
\eeqa
These can be extracted from the {generating function} 
\beqa
\sum_{b=0}^\infty \mathcal{J}^b_\alpha \eta^b&=&\!\int\limits_0^\Delta \! \frac{\epsilon^\alpha d\epsilon}{1-\eta \frac{1-\epsilon}{1+\epsilon}} \nonumber\\&=&\frac{\Delta^{\alpha+1}}{\alpha+1}\frac{\eta-1-2\eta \,_2 F_1(1,\alpha+1,\alpha+2,\Delta \frac{\eta+1}{\eta-1})}{\eta^2-1} \,. \nn
\eeqa
%
Expanding this expression for large $\Delta$ first and taking the $b$-th series coefficient yields the desired expansion for large $\Delta$ of the integrals $\mathcal{J}^{b}_\alpha(\Delta)$.

At the end of the day, adding up inner and outer conditions we observe as before all cut-offs nicely dropping out and we obtain in this way a final expansion of the form
\beq
J_n^{bc}(s)=\sum_{j=0}^{14}\frac{e_j^n(b,c)+\log(s)\,f_j^n(b,c)}{s^{j/2}} +\mathcal{O}(s^{-15/2})
\eeq
which we can then reassemble into the desired expansion~(\ref{goalExpansion}) using (\ref{measureDecomp}).

\section{Fits in $L$ and in $N $} \label{numDet}
In a primal numeric formulation we construct solutions which should approach the optimal solution as all numerical truncation parameters (number of spins, number of grid points, number of parameters in the ansatz) tend to infinity. In practice we need to keep them finite and extrapolate and that always involves some arbitrariness. Here we describe how we extrapolated each of our numerics to $L  \to \infty$ first for each value of $N $ and then how these   results were extrapolated further to obtain an estimate of the optimal bound for $N  \to \infty$. (We repeated this exercise for two different grids in $s$ and found nearly identical results so we believe the~$L$ and~$N$ fits which we are about to describe in detail are really the most relevant ones.)

\begin{figure}[t]
\centering
        \includegraphics[scale=0.39]{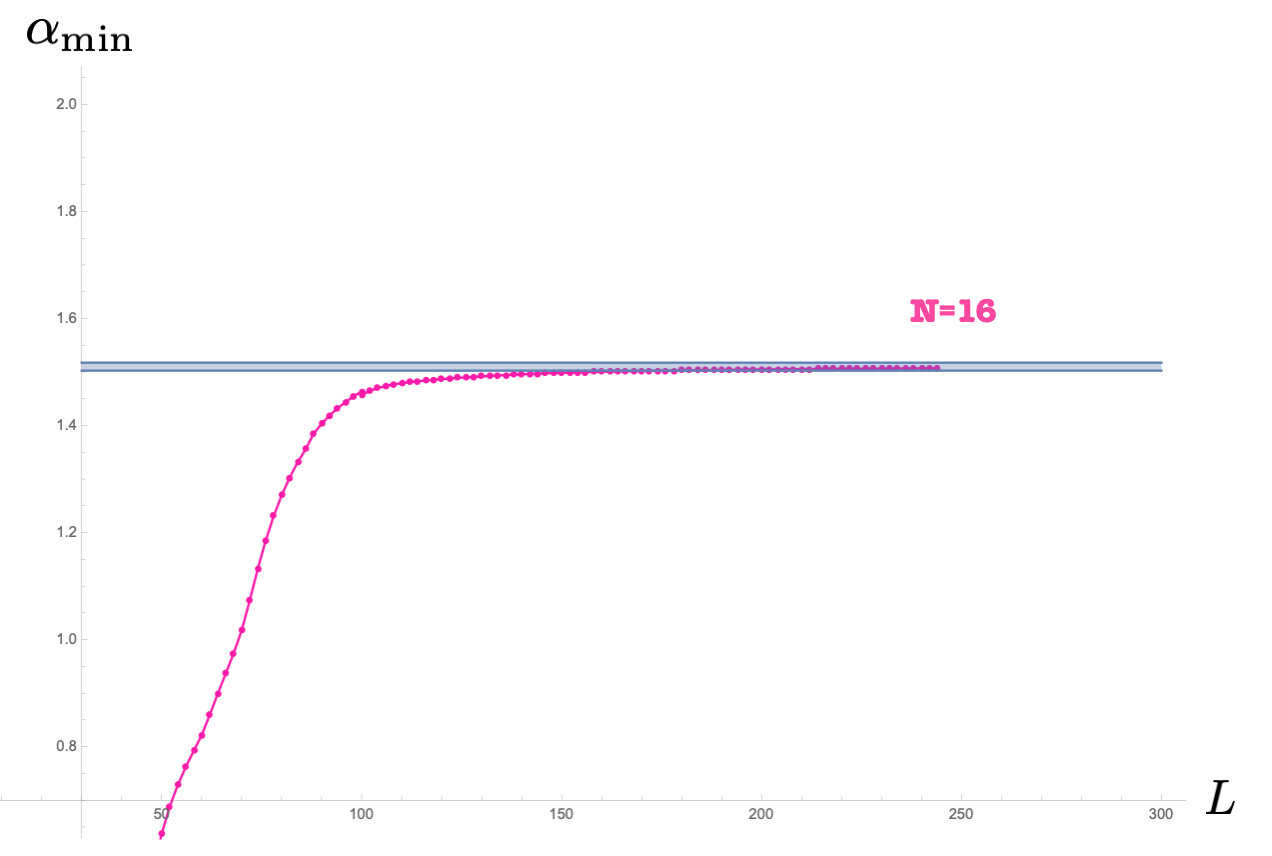}
\caption{Plot of $\alpha^\text{min}$ as a function of $L$ for fixed $N=16$. The blue region is the estimate for  $L\to \infty$ as described in the text.} 
\label{exampleFixedN}
\end{figure}

First let us describe the large $L$ fit using figure \ref{exampleFixedN} as an example. In this figure the various dots are the minimum value of $\alpha$ attained by our ansatz with $N=16$ and imposing unitarity for all spins up to $L  $ which is varied from $40$ to $244$. 

To extrapolate these numerics to $L  = \infty$ we consider a window including the last $m$ points. (Soon we will scan over $m$.) Then we try to fit these last $m$ points using a variety of ansatze such as 
\beqa
a+b/L \,, \,\,\,\,a+b/L+c/L^2 \,, \,\,\,\, a+b/L^c \, ,\,\,\,\, a+b e^{-c L}  \, \dots  \nonumber
\eeqa
In all these ansatz the estimate for the $L=\infty$ asymptote is $a$.\footnote{We drop all fits where $a$ is smaller than the last~$L  =244$ data point since the curve ought to be monotonic after all; we also drop ansatze which predict too large $a$ which are also clearly wrong.}
We repeat this for a large number of $m$ points.\footnote{We take $m$ to be large enough so that we are not overfitting but not so large that   we are capturing more than the  asymptotic region.} In this way we generate a large number of fits $\texttt{fit}^{(n)}$. Each of them gives an estimate for $a=a^{(n)}$ and comes with its own weight function, \footnote{The choice of the weight function is arbitrary: our definition resembles a $\chi^2$. We have checked the stability of our estimate by choosing different reasonable functions and obtaining results compatible within the errors. } 
\beq
(\chi^{(n)})^2 = \frac{1}{m} \sum_{j=1}^m (\texttt{fit$^{(n)}$ estimate}_j -\texttt{data point}_j)^2 \nonumber
\eeq
Finally, we estimate the optimal value of $a$ by averaging all these fits weighing them according to how well they manage to fit the numerics, 
\beqa
a_\texttt{estimate}=\frac{\sum_n a^{(n)} /(\chi^{(n)})^2 }{\sum_n 1 /(\chi^{(n)})^2}\nonumber
\eeqa
We also estimate the variance in a similar way 
\beq
\sigma^2  = \frac{\sum_n (a^{(n)}-a_\texttt{estimate} )^2 /(\chi^{(n)})^2 }{\sum_n 1 /(\chi^{(n)})^2}
\eeq
For $N=16$ in the figure we obtain in this way an estimate for $a=1.51 \pm 0.01$ represented by the blue strip in figure \ref{exampleFixedN}. We now repeat this for all $N$'s. That is how we produced the dots and corresponding error bars in figure~\ref{AllNs}.

Finally, we fitted these points together with the corresponding error bars with an ansatz $A+B/N^C$ using \texttt{Mathematica}'s built in function \texttt{NonlinearModelFit} which yields an estimate for $N=\infty$ (i.e. $A$) together with its associated error. We got in this way 
\beq
A=0.13 \pm 0.02\nonumber
\eeq
which is what is represented in the green strip in figure~\ref{AllNs}. 

The data used for these fits (\texttt{data.txt}) and a notebook leading to these estimates (\texttt{fits.nb})  -- and  containing details about all fits and windows mentioned above -- is attached to this submission. The reader is encouraged to try other (hopefully more sophisticated) statistical analysis. Comments and suggestions in this regard would be most welcome.

\section{Inverse Amplitude Method}
\label{ap:IAM}

The inverse amplitude method (IAM) is a technique to build approximate partial amplitudes using elastic unitarity and perturbation theory \cite{Truong:1988zp, Dobado:1996ps}.
Let us briefly review the IAM in its simplest form. 
The first step is to write 
\beq
S_\ell(s) = 1+i t_\ell(s)\,,
\eeq
so that the elastic unitarity condition $|S_\ell| = 1$ becomes
\beq
2{\rm Im}\, t_\ell = |t_\ell|^2\,.
\eeq
The main observation is to note that this implies
\beq
2{\rm Im}\,\frac{1}{ t_\ell} =  -1\,.
\eeq
Therefore, an $n$-loop perturbative expansion for $1/t_\ell$ will obey elastic unitarity exactly for any $n\ge 1$.

In our case, using the results of appendix \ref{unitAppendix}, we obtain the following leading order IAM approximation 
\beq
\frac{1}{t_\ell(s)} \approx \frac{1}{2 \delta_\ell^{(0)} (\ell_P^2 s)^4} - \frac{i}{2}\,,
\eeq
or equivalently
\beq
S_\ell(s) \approx \frac{i - \delta_\ell^{(0)} (\ell_P^2 s)^4 }{i + \delta_\ell^{(0)} (\ell_P^2 s)^4 } \,.
\label{SlIAM}
\eeq
This approximation gives rise to resonances at
\beq
 s= m^2(\ell) = \frac{1}{\ell_P^2} e^{i\pi/8} \left[ \delta_\ell^{(0)} \right]^{-\frac{1}{4}}\,.
\eeq
In particular, the spin zero resonance (or graviball) has $\ell_P^2 m^2 (\ell=0)\approx
3.4 +  1.4\, i $, which has a similar real part to the lightest resonance found in our numerical approach (see figure \ref{spin0resonance}).
The large spin scaling $m^2 \sim \ell^\frac{3}{2}$ also agrees with our numerical results.
Finally, using 
\beq
{\rm Im} \,T(s,t=0) = 2 s^{-3} \sum_{\ell=0 \atop even}^\infty \left(1-{\rm Re}\,S_\ell(s)\right)
P_\ell^{(d)}\left(1\right)\,,
\label{PWEimT}
\eeq
with the IAM approximation \eqref{SlIAM}, we can estimate the integrand of the sum rule \eqref{alphasumrule}. This is shown in blue in figure \ref{ImTintegral} and it gives $\alpha \approx 0.7$, which violates our numerical lower bound. This is not surprising because the IAM does not respect crossing nor analyticity. In fact, the approximation \eqref{SlIAM} has unacceptable poles at complex locations on the physical sheet. Nevertheless, the IAM is usually a good starting point to get a qualitative idea of how the non-perturbative amplitudes may look like.

\bibliography{WhereIsStringTheory_v01Notes} 

\begin{thebibliography}{50}%
\makeatletter
\providecommand \@ifxundefined [1]{%
 \@ifx{#1\undefined}
}%
\providecommand \@ifnum [1]{%
 \ifnum #1\expandafter \@firstoftwo
 \else \expandafter \@secondoftwo
 \fi
}%
\providecommand \@ifx [1]{%
 \ifx #1\expandafter \@firstoftwo
 \else \expandafter \@secondoftwo
 \fi
}%
\providecommand \natexlab [1]{#1}%
\providecommand \enquote  [1]{``#1''}%
\providecommand \bibnamefont  [1]{#1}%
\providecommand \bibfnamefont [1]{#1}%
\providecommand \citenamefont [1]{#1}%
\providecommand \href@noop [0]{\@secondoftwo}%
\providecommand \href [0]{\begingroup \@sanitize@url \@href}%
\providecommand \@href[1]{\@@startlink{#1}\@@href}%
\providecommand \@@href[1]{\endgroup#1\@@endlink}%
\providecommand \@sanitize@url [0]{\catcode `\\12\catcode `\$12\catcode
  `\&12\catcode `\#12\catcode `\^12\catcode `\_12\catcode `\%12\relax}%
\providecommand \@@startlink[1]{}%
\providecommand \@@endlink[0]{}%
\providecommand \url  [0]{\begingroup\@sanitize@url \@url }%
\providecommand \@url [1]{\endgroup\@href {#1}{\urlprefix }}%
\providecommand \urlprefix  [0]{URL }%
\providecommand \Eprint [0]{\href }%
\providecommand \doibase [0]{http://dx.doi.org/}%
\providecommand \selectlanguage [0]{\@gobble}%
\providecommand \bibinfo  [0]{\@secondoftwo}%
\providecommand \bibfield  [0]{\@secondoftwo}%
\providecommand \translation [1]{[#1]}%
\providecommand \BibitemOpen [0]{}%
\providecommand \bibitemStop [0]{}%
\providecommand \bibitemNoStop [0]{.\EOS\space}%
\providecommand \EOS [0]{\spacefactor3000\relax}%
\providecommand \BibitemShut  [1]{\csname bibitem#1\endcsname}%
\let\auto@bib@innerbib\@empty
\bibitem [{\citenamefont {Elias~Mir\'o}\ \emph {et~al.}(2019)\citenamefont
  {Elias~Mir\'o}, \citenamefont {Guerrieri}, \citenamefont {Hebbar},
  \citenamefont {Penedones},\ and\ \citenamefont {Vieira}}]{FluxTube}%
  \BibitemOpen
  \bibfield  {author} {\bibinfo {author} {\bibfnamefont {Joan}\ \bibnamefont
  {Elias~Mir\'o}}, \bibinfo {author} {\bibfnamefont {Andrea~L.}\ \bibnamefont
  {Guerrieri}}, \bibinfo {author} {\bibfnamefont {Aditya}\ \bibnamefont
  {Hebbar}}, \bibinfo {author} {\bibfnamefont {Joao}\ \bibnamefont
  {Penedones}}, \ and\ \bibinfo {author} {\bibfnamefont {Pedro}\ \bibnamefont
  {Vieira}},\ }\bibfield  {title} {\enquote {\bibinfo {title} {{Flux Tube
  S-matrix Bootstrap}},}\ }\href {\doibase 10.1103/PhysRevLett.123.221602}
  {\bibfield  {journal} {\bibinfo  {journal} {Phys. Rev. Lett.}\ }\textbf
  {\bibinfo {volume} {123}},\ \bibinfo {pages} {221602} (\bibinfo {year}
  {2019})},\ \Eprint {http://arxiv.org/abs/1906.08098} {arXiv:1906.08098
  [hep-th]} \BibitemShut {NoStop}%
\bibitem [{\citenamefont {Guerrieri}\ \emph
  {et~al.}(2020{\natexlab{a}})\citenamefont {Guerrieri}, \citenamefont
  {Penedones},\ and\ \citenamefont {Vieira}}]{Pions2}%
  \BibitemOpen
  \bibfield  {author} {\bibinfo {author} {\bibfnamefont {Andrea}\ \bibnamefont
  {Guerrieri}}, \bibinfo {author} {\bibfnamefont {Joao}\ \bibnamefont
  {Penedones}}, \ and\ \bibinfo {author} {\bibfnamefont {Pedro}\ \bibnamefont
  {Vieira}},\ }\bibfield  {title} {\enquote {\bibinfo {title} {{S-matrix
  Bootstrap for Effective Field Theories: Massless Pions}},}\ }\href@noop {} {\
   (\bibinfo {year} {2020}{\natexlab{a}})},\ \Eprint
  {http://arxiv.org/abs/2011.02802} {arXiv:2011.02802 [hep-th]} \BibitemShut
  {NoStop}%
\bibitem [{\citenamefont {Green}\ \emph {et~al.}()\citenamefont {Green},
  \citenamefont {Schwarz},\ and\ \citenamefont {Witten}}]{GSW}%
  \BibitemOpen
  \bibfield  {author} {\bibinfo {author} {\bibfnamefont {M.~B.}\ \bibnamefont
  {Green}}, \bibinfo {author} {\bibfnamefont {J.~H.}\ \bibnamefont {Schwarz}},
  \ and\ \bibinfo {author} {\bibfnamefont {E.}~\bibnamefont {Witten}},\
  }\href@noop {} {\bibinfo  {journal} {{Superstring Theory Vol. 1:
  Introduction, Cambridge Univ. Press (1987) 469 p. (Cambridge Monographs On
  Mathematical Physics)}}\ }\BibitemShut {NoStop}%
\bibitem [{\citenamefont {Alday}\ and\ \citenamefont {Maldacena}(2007)}]{juan}%
  \BibitemOpen
\bibfield  {journal} {  }\bibfield  {author} {\bibinfo {author} {\bibfnamefont
  {Luis~F.}\ \bibnamefont {Alday}}\ and\ \bibinfo {author} {\bibfnamefont
  {Juan~Martin}\ \bibnamefont {Maldacena}},\ }\bibfield  {title} {\enquote
  {\bibinfo {title} {{Gluon scattering amplitudes at strong coupling}},}\
  }\href {\doibase 10.1088/1126-6708/2007/06/064} {\bibfield  {journal}
  {\bibinfo  {journal} {JHEP}\ }\textbf {\bibinfo {volume} {06}},\ \bibinfo
  {pages} {064} (\bibinfo {year} {2007})},\ \Eprint
  {http://arxiv.org/abs/0705.0303} {arXiv:0705.0303 [hep-th]} \BibitemShut
  {NoStop}%
\bibitem [{\citenamefont {Boels}\ and\ \citenamefont
  {O'Connell}(2012)}]{Boels:2012ie}%
  \BibitemOpen
  \bibfield  {author} {\bibinfo {author} {\bibfnamefont {Rutger~H.}\
  \bibnamefont {Boels}}\ and\ \bibinfo {author} {\bibfnamefont {Donal}\
  \bibnamefont {O'Connell}},\ }\bibfield  {title} {\enquote {\bibinfo {title}
  {{Simple superamplitudes in higher dimensions}},}\ }\href {\doibase
  10.1007/JHEP06(2012)163} {\bibfield  {journal} {\bibinfo  {journal} {JHEP}\
  }\textbf {\bibinfo {volume} {06}},\ \bibinfo {pages} {163} (\bibinfo {year}
  {2012})},\ \Eprint {http://arxiv.org/abs/1201.2653} {arXiv:1201.2653
  [hep-th]} \BibitemShut {NoStop}%
\bibitem [{\citenamefont {Wang}\ and\ \citenamefont {Yin}(2015)}]{XiYinPaper}%
  \BibitemOpen
  \bibfield  {author} {\bibinfo {author} {\bibfnamefont {Yifan}\ \bibnamefont
  {Wang}}\ and\ \bibinfo {author} {\bibfnamefont {Xi}~\bibnamefont {Yin}},\
  }\bibfield  {title} {\enquote {\bibinfo {title} {{Constraining Higher
  Derivative Supergravity with Scattering Amplitudes}},}\ }\href {\doibase
  10.1103/PhysRevD.92.041701} {\bibfield  {journal} {\bibinfo  {journal} {Phys.
  Rev. D}\ }\textbf {\bibinfo {volume} {92}},\ \bibinfo {pages} {041701}
  (\bibinfo {year} {2015})},\ \Eprint {http://arxiv.org/abs/1502.03810}
  {arXiv:1502.03810 [hep-th]} \BibitemShut {NoStop}%
\bibitem [{\citenamefont {Policastro}\ and\ \citenamefont
  {Tsimpis}(2006)}]{R4PureSpinor}%
  \BibitemOpen
  \bibfield  {author} {\bibinfo {author} {\bibfnamefont {Giuseppe}\
  \bibnamefont {Policastro}}\ and\ \bibinfo {author} {\bibfnamefont
  {Dimitrios}\ \bibnamefont {Tsimpis}},\ }\bibfield  {title} {\enquote
  {\bibinfo {title} {{R**4, purified}},}\ }\href {\doibase
  10.1088/0264-9381/23/14/012} {\bibfield  {journal} {\bibinfo  {journal}
  {Class. Quant. Grav.}\ }\textbf {\bibinfo {volume} {23}},\ \bibinfo {pages}
  {4753--4780} (\bibinfo {year} {2006})},\ \Eprint
  {http://arxiv.org/abs/hep-th/0603165} {arXiv:hep-th/0603165} \BibitemShut
  {NoStop}%
\bibitem [{\citenamefont {Gross}\ and\ \citenamefont
  {Witten}(1986)}]{Gross:1986iv}%
  \BibitemOpen
  \bibfield  {author} {\bibinfo {author} {\bibfnamefont {David~J.}\
  \bibnamefont {Gross}}\ and\ \bibinfo {author} {\bibfnamefont {Edward}\
  \bibnamefont {Witten}},\ }\bibfield  {title} {\enquote {\bibinfo {title}
  {{Superstring Modifications of Einstein's Equations}},}\ }\href {\doibase
  10.1016/0550-3213(86)90429-3} {\bibfield  {journal} {\bibinfo  {journal}
  {Nucl. Phys. B}\ }\textbf {\bibinfo {volume} {277}},\ \bibinfo {pages} {1}
  (\bibinfo {year} {1986})}\BibitemShut {NoStop}%
\bibitem [{\citenamefont {Green}\ and\ \citenamefont
  {Gutperle}(1997)}]{Green:1997tv}%
  \BibitemOpen
  \bibfield  {author} {\bibinfo {author} {\bibfnamefont {Michael~B.}\
  \bibnamefont {Green}}\ and\ \bibinfo {author} {\bibfnamefont {Michael}\
  \bibnamefont {Gutperle}},\ }\bibfield  {title} {\enquote {\bibinfo {title}
  {{Effects of D instantons}},}\ }\href {\doibase
  10.1016/S0550-3213(97)00269-1} {\bibfield  {journal} {\bibinfo  {journal}
  {Nucl. Phys. B}\ }\textbf {\bibinfo {volume} {498}},\ \bibinfo {pages}
  {195--227} (\bibinfo {year} {1997})},\ \Eprint
  {http://arxiv.org/abs/hep-th/9701093} {arXiv:hep-th/9701093} \BibitemShut
  {NoStop}%
\bibitem [{\citenamefont {Green}\ and\ \citenamefont
  {Vanhove}(1997)}]{Green:1997di}%
  \BibitemOpen
  \bibfield  {author} {\bibinfo {author} {\bibfnamefont {Michael~B.}\
  \bibnamefont {Green}}\ and\ \bibinfo {author} {\bibfnamefont {Pierre}\
  \bibnamefont {Vanhove}},\ }\bibfield  {title} {\enquote {\bibinfo {title} {{D
  instantons, strings and M theory}},}\ }\href {\doibase
  10.1016/S0370-2693(97)00785-5} {\bibfield  {journal} {\bibinfo  {journal}
  {Phys. Lett. B}\ }\textbf {\bibinfo {volume} {408}},\ \bibinfo {pages}
  {122--134} (\bibinfo {year} {1997})},\ \Eprint
  {http://arxiv.org/abs/hep-th/9704145} {arXiv:hep-th/9704145} \BibitemShut
  {NoStop}%
\bibitem [{\citenamefont {Green}\ \emph {et~al.}(2007)\citenamefont {Green},
  \citenamefont {Russo},\ and\ \citenamefont {Vanhove}}]{Green:2006gt}%
  \BibitemOpen
  \bibfield  {author} {\bibinfo {author} {\bibfnamefont {Michael~B.}\
  \bibnamefont {Green}}, \bibinfo {author} {\bibfnamefont {Jorge~G.}\
  \bibnamefont {Russo}}, \ and\ \bibinfo {author} {\bibfnamefont {Pierre}\
  \bibnamefont {Vanhove}},\ }\bibfield  {title} {\enquote {\bibinfo {title}
  {{Non-renormalisation conditions in type II string theory and maximal
  supergravity}},}\ }\href {\doibase 10.1088/1126-6708/2007/02/099} {\bibfield
  {journal} {\bibinfo  {journal} {JHEP}\ }\textbf {\bibinfo {volume} {02}},\
  \bibinfo {pages} {099} (\bibinfo {year} {2007})},\ \Eprint
  {http://arxiv.org/abs/hep-th/0610299} {arXiv:hep-th/0610299} \BibitemShut
  {NoStop}%
\bibitem [{\citenamefont {Chester}\ \emph
  {et~al.}(2020{\natexlab{a}})\citenamefont {Chester}, \citenamefont {Green},
  \citenamefont {Pufu}, \citenamefont {Wang},\ and\ \citenamefont
  {Wen}}]{Chester:2019jas}%
  \BibitemOpen
  \bibfield  {author} {\bibinfo {author} {\bibfnamefont {Shai~M.}\ \bibnamefont
  {Chester}}, \bibinfo {author} {\bibfnamefont {Michael~B.}\ \bibnamefont
  {Green}}, \bibinfo {author} {\bibfnamefont {Silviu~S.}\ \bibnamefont {Pufu}},
  \bibinfo {author} {\bibfnamefont {Yifan}\ \bibnamefont {Wang}}, \ and\
  \bibinfo {author} {\bibfnamefont {Congkao}\ \bibnamefont {Wen}},\ }\bibfield
  {title} {\enquote {\bibinfo {title} {{Modular invariance in superstring
  theory from $ \mathcal{N} $ = 4 super-Yang-Mills}},}\ }\href {\doibase
  10.1007/JHEP11(2020)016} {\bibfield  {journal} {\bibinfo  {journal} {JHEP}\
  }\textbf {\bibinfo {volume} {11}},\ \bibinfo {pages} {016} (\bibinfo {year}
  {2020}{\natexlab{a}})},\ \Eprint {http://arxiv.org/abs/1912.13365}
  {arXiv:1912.13365 [hep-th]} \BibitemShut {NoStop}%
\bibitem [{\citenamefont {Pioline}(2015)}]{Pioline:2015yea}%
  \BibitemOpen
  \bibfield  {author} {\bibinfo {author} {\bibfnamefont {Boris}\ \bibnamefont
  {Pioline}},\ }\bibfield  {title} {\enquote {\bibinfo {title}
  {{D$^{6}${R}$^{4}$ amplitudes in various dimensions}},}\ }\href {\doibase
  10.1007/JHEP04(2015)057} {\bibfield  {journal} {\bibinfo  {journal} {JHEP}\
  }\textbf {\bibinfo {volume} {04}},\ \bibinfo {pages} {057} (\bibinfo {year}
  {2015})},\ \Eprint {http://arxiv.org/abs/1502.03377} {arXiv:1502.03377
  [hep-th]} \BibitemShut {NoStop}%
\bibitem [{\citenamefont {Binder}\ \emph {et~al.}(2020)\citenamefont {Binder},
  \citenamefont {Chester},\ and\ \citenamefont {Pufu}}]{Binder:2019mpb}%
  \BibitemOpen
  \bibfield  {author} {\bibinfo {author} {\bibfnamefont {Damon~J.}\
  \bibnamefont {Binder}}, \bibinfo {author} {\bibfnamefont {Shai~M.}\
  \bibnamefont {Chester}}, \ and\ \bibinfo {author} {\bibfnamefont {Silviu~S.}\
  \bibnamefont {Pufu}},\ }\bibfield  {title} {\enquote {\bibinfo {title}
  {{AdS$_{4}$/CFT$_{3}$ from weak to strong string coupling}},}\ }\href
  {\doibase 10.1007/JHEP01(2020)034} {\bibfield  {journal} {\bibinfo  {journal}
  {JHEP}\ }\textbf {\bibinfo {volume} {01}},\ \bibinfo {pages} {034} (\bibinfo
  {year} {2020})},\ \Eprint {http://arxiv.org/abs/1906.07195} {arXiv:1906.07195
  [hep-th]} \BibitemShut {NoStop}%
\bibitem [{\citenamefont {Adams}\ \emph {et~al.}(2006)\citenamefont {Adams},
  \citenamefont {Arkani-Hamed}, \citenamefont {Dubovsky}, \citenamefont
  {Nicolis},\ and\ \citenamefont {Rattazzi}}]{NimaEtAl}%
  \BibitemOpen
  \bibfield  {author} {\bibinfo {author} {\bibfnamefont {Allan}\ \bibnamefont
  {Adams}}, \bibinfo {author} {\bibfnamefont {Nima}\ \bibnamefont
  {Arkani-Hamed}}, \bibinfo {author} {\bibfnamefont {Sergei}\ \bibnamefont
  {Dubovsky}}, \bibinfo {author} {\bibfnamefont {Alberto}\ \bibnamefont
  {Nicolis}}, \ and\ \bibinfo {author} {\bibfnamefont {Riccardo}\ \bibnamefont
  {Rattazzi}},\ }\bibfield  {title} {\enquote {\bibinfo {title} {{Causality,
  analyticity and an IR obstruction to UV completion}},}\ }\href {\doibase
  10.1088/1126-6708/2006/10/014} {\bibfield  {journal} {\bibinfo  {journal}
  {JHEP}\ }\textbf {\bibinfo {volume} {10}},\ \bibinfo {pages} {014} (\bibinfo
  {year} {2006})},\ \Eprint {http://arxiv.org/abs/hep-th/0602178}
  {arXiv:hep-th/0602178} \BibitemShut {NoStop}%
\bibitem [{\citenamefont {Arkani-Hamed}\ \emph {et~al.}(2020)\citenamefont
  {Arkani-Hamed}, \citenamefont {Huang},\ and\ \citenamefont
  {Huang}}]{Arkanihedron}%
  \BibitemOpen
  \bibfield  {author} {\bibinfo {author} {\bibfnamefont {Nima}\ \bibnamefont
  {Arkani-Hamed}}, \bibinfo {author} {\bibfnamefont {Tzu-Chen}\ \bibnamefont
  {Huang}}, \ and\ \bibinfo {author} {\bibfnamefont {Yu-Tin}\ \bibnamefont
  {Huang}},\ }\bibfield  {title} {\enquote {\bibinfo {title} {{The
  EFT-Hedron}},}\ }\href@noop {} {\  (\bibinfo {year} {2020})},\ \Eprint
  {http://arxiv.org/abs/2012.15849} {arXiv:2012.15849 [hep-th]} \BibitemShut
  {NoStop}%
\bibitem [{\citenamefont {Bellazzini}\ \emph {et~al.}(2020)\citenamefont
  {Bellazzini}, \citenamefont {Elias~Mir\'o}, \citenamefont {Rattazzi},
  \citenamefont {Riembau},\ and\ \citenamefont {Riva}}]{Rattazzipositive}%
  \BibitemOpen
  \bibfield  {author} {\bibinfo {author} {\bibfnamefont {Brando}\ \bibnamefont
  {Bellazzini}}, \bibinfo {author} {\bibfnamefont {Joan}\ \bibnamefont
  {Elias~Mir\'o}}, \bibinfo {author} {\bibfnamefont {Riccardo}\ \bibnamefont
  {Rattazzi}}, \bibinfo {author} {\bibfnamefont {Marc}\ \bibnamefont
  {Riembau}}, \ and\ \bibinfo {author} {\bibfnamefont {Francesco}\ \bibnamefont
  {Riva}},\ }\bibfield  {title} {\enquote {\bibinfo {title} {{Positive Moments
  for Scattering Amplitudes}},}\ }\href@noop {} {\  (\bibinfo {year} {2020})},\
  \Eprint {http://arxiv.org/abs/2011.00037} {arXiv:2011.00037 [hep-th]}
  \BibitemShut {NoStop}%
\bibitem [{\citenamefont {Tolley}\ \emph {et~al.}(2020)\citenamefont {Tolley},
  \citenamefont {Wang},\ and\ \citenamefont {Zhou}}]{Tolley:2020gtv}%
  \BibitemOpen
  \bibfield  {author} {\bibinfo {author} {\bibfnamefont {Andrew~J.}\
  \bibnamefont {Tolley}}, \bibinfo {author} {\bibfnamefont {Zi-Yue}\
  \bibnamefont {Wang}}, \ and\ \bibinfo {author} {\bibfnamefont {Shuang-Yong}\
  \bibnamefont {Zhou}},\ }\bibfield  {title} {\enquote {\bibinfo {title} {{New
  positivity bounds from full crossing symmetry}},}\ }\href@noop {} {\
  (\bibinfo {year} {2020})},\ \Eprint {http://arxiv.org/abs/2011.02400}
  {arXiv:2011.02400 [hep-th]} \BibitemShut {NoStop}%
\bibitem [{\citenamefont {Caron-Huot}\ and\ \citenamefont
  {Van~Duong}(2020)}]{simoneft}%
  \BibitemOpen
  \bibfield  {author} {\bibinfo {author} {\bibfnamefont {Simon}\ \bibnamefont
  {Caron-Huot}}\ and\ \bibinfo {author} {\bibfnamefont {Vincent}\ \bibnamefont
  {Van~Duong}},\ }\bibfield  {title} {\enquote {\bibinfo {title} {{Extremal
  Effective Field Theories}},}\ }\href@noop {} {\  (\bibinfo {year} {2020})},\
  \Eprint {http://arxiv.org/abs/2011.02957} {arXiv:2011.02957 [hep-th]}
  \BibitemShut {NoStop}%
\bibitem [{\citenamefont {Paulos}\ \emph {et~al.}(2019)\citenamefont {Paulos},
  \citenamefont {Penedones}, \citenamefont {Toledo}, \citenamefont {van Rees},\
  and\ \citenamefont {Vieira}}]{4dpaper1}%
  \BibitemOpen
  \bibfield  {author} {\bibinfo {author} {\bibfnamefont {Miguel~F.}\
  \bibnamefont {Paulos}}, \bibinfo {author} {\bibfnamefont {Joao}\ \bibnamefont
  {Penedones}}, \bibinfo {author} {\bibfnamefont {Jonathan}\ \bibnamefont
  {Toledo}}, \bibinfo {author} {\bibfnamefont {Balt~C.}\ \bibnamefont {van
  Rees}}, \ and\ \bibinfo {author} {\bibfnamefont {Pedro}\ \bibnamefont
  {Vieira}},\ }\bibfield  {title} {\enquote {\bibinfo {title} {{The S-matrix
  bootstrap. Part III: higher dimensional amplitudes}},}\ }\href {\doibase
  10.1007/JHEP12(2019)040} {\bibfield  {journal} {\bibinfo  {journal} {JHEP}\
  }\textbf {\bibinfo {volume} {12}},\ \bibinfo {pages} {040} (\bibinfo {year}
  {2019})},\ \Eprint {http://arxiv.org/abs/1708.06765} {arXiv:1708.06765
  [hep-th]} \BibitemShut {NoStop}%
\bibitem [{\citenamefont {Guerrieri}\ \emph {et~al.}(2019)\citenamefont
  {Guerrieri}, \citenamefont {Penedones},\ and\ \citenamefont
  {Vieira}}]{Pions1}%
  \BibitemOpen
  \bibfield  {author} {\bibinfo {author} {\bibfnamefont {Andrea~L.}\
  \bibnamefont {Guerrieri}}, \bibinfo {author} {\bibfnamefont {Joao}\
  \bibnamefont {Penedones}}, \ and\ \bibinfo {author} {\bibfnamefont {Pedro}\
  \bibnamefont {Vieira}},\ }\bibfield  {title} {\enquote {\bibinfo {title}
  {{Bootstrapping QCD Using Pion Scattering Amplitudes}},}\ }\href {\doibase
  10.1103/PhysRevLett.122.241604} {\bibfield  {journal} {\bibinfo  {journal}
  {Phys. Rev. Lett.}\ }\textbf {\bibinfo {volume} {122}},\ \bibinfo {pages}
  {241604} (\bibinfo {year} {2019})},\ \Eprint
  {http://arxiv.org/abs/1810.12849} {arXiv:1810.12849 [hep-th]} \BibitemShut
  {NoStop}%
\bibitem [{\citenamefont {Hebbar}\ \emph {et~al.}(2020)\citenamefont {Hebbar},
  \citenamefont {Karateev},\ and\ \citenamefont {Penedones}}]{4dspinning}%
  \BibitemOpen
  \bibfield  {author} {\bibinfo {author} {\bibfnamefont {Aditya}\ \bibnamefont
  {Hebbar}}, \bibinfo {author} {\bibfnamefont {Denis}\ \bibnamefont
  {Karateev}}, \ and\ \bibinfo {author} {\bibfnamefont {Joao}\ \bibnamefont
  {Penedones}},\ }\bibfield  {title} {\enquote {\bibinfo {title} {{Spinning
  S-matrix Bootstrap in 4d}},}\ }\href@noop {} {\  (\bibinfo {year} {2020})},\
  \Eprint {http://arxiv.org/abs/2011.11708} {arXiv:2011.11708 [hep-th]}
  \BibitemShut {NoStop}%
\bibitem [{\citenamefont {Bose}\ \emph
  {et~al.}(2020{\natexlab{a}})\citenamefont {Bose}, \citenamefont {Haldar},
  \citenamefont {Sinha}, \citenamefont {Sinha},\ and\ \citenamefont
  {Tiwari}}]{aninda1}%
  \BibitemOpen
  \bibfield  {author} {\bibinfo {author} {\bibfnamefont {Anjishnu}\
  \bibnamefont {Bose}}, \bibinfo {author} {\bibfnamefont {Parthiv}\
  \bibnamefont {Haldar}}, \bibinfo {author} {\bibfnamefont {Aninda}\
  \bibnamefont {Sinha}}, \bibinfo {author} {\bibfnamefont {Pritish}\
  \bibnamefont {Sinha}}, \ and\ \bibinfo {author} {\bibfnamefont {Shaswat~S.}\
  \bibnamefont {Tiwari}},\ }\bibfield  {title} {\enquote {\bibinfo {title}
  {{Relative entropy in scattering and the S-matrix bootstrap}},}\ }\href
  {\doibase 10.21468/SciPostPhys.9.5.081} {\bibfield  {journal} {\bibinfo
  {journal} {SciPost Phys.}\ }\textbf {\bibinfo {volume} {9}},\ \bibinfo
  {pages} {081} (\bibinfo {year} {2020}{\natexlab{a}})},\ \Eprint
  {http://arxiv.org/abs/2006.12213} {arXiv:2006.12213 [hep-th]} \BibitemShut
  {NoStop}%
\bibitem [{\citenamefont {Bose}\ \emph
  {et~al.}(2020{\natexlab{b}})\citenamefont {Bose}, \citenamefont {Sinha},\
  and\ \citenamefont {Tiwari}}]{aninda2}%
  \BibitemOpen
  \bibfield  {author} {\bibinfo {author} {\bibfnamefont {Anjishnu}\
  \bibnamefont {Bose}}, \bibinfo {author} {\bibfnamefont {Aninda}\ \bibnamefont
  {Sinha}}, \ and\ \bibinfo {author} {\bibfnamefont {Shaswat~S.}\ \bibnamefont
  {Tiwari}},\ }\bibfield  {title} {\enquote {\bibinfo {title} {{Selection rules
  for the S-Matrix bootstrap}},}\ }\href@noop {} {\  (\bibinfo {year}
  {2020}{\natexlab{b}})},\ \Eprint {http://arxiv.org/abs/2011.07944}
  {arXiv:2011.07944 [hep-th]} \BibitemShut {NoStop}%
\bibitem [{\citenamefont {C\'ordova}\ \emph {et~al.}(2020)\citenamefont
  {C\'ordova}, \citenamefont {He}, \citenamefont {Kruczenski},\ and\
  \citenamefont {Vieira}}]{monolith}%
  \BibitemOpen
  \bibfield  {author} {\bibinfo {author} {\bibfnamefont {Luc\'\i{}a}\
  \bibnamefont {C\'ordova}}, \bibinfo {author} {\bibfnamefont {Yifei}\
  \bibnamefont {He}}, \bibinfo {author} {\bibfnamefont {Martin}\ \bibnamefont
  {Kruczenski}}, \ and\ \bibinfo {author} {\bibfnamefont {Pedro}\ \bibnamefont
  {Vieira}},\ }\bibfield  {title} {\enquote {\bibinfo {title} {{The O(N)
  S-matrix Monolith}},}\ }\href {\doibase 10.1007/JHEP04(2020)142} {\bibfield
  {journal} {\bibinfo  {journal} {JHEP}\ }\textbf {\bibinfo {volume} {04}},\
  \bibinfo {pages} {142} (\bibinfo {year} {2020})},\ \Eprint
  {http://arxiv.org/abs/1909.06495} {arXiv:1909.06495 [hep-th]} \BibitemShut
  {NoStop}%
\bibitem [{\citenamefont {Guerrieri}\ \emph
  {et~al.}(2020{\natexlab{b}})\citenamefont {Guerrieri}, \citenamefont
  {Homrich},\ and\ \citenamefont {Vieira}}]{2dpaper}%
  \BibitemOpen
  \bibfield  {author} {\bibinfo {author} {\bibfnamefont {Andrea~L.}\
  \bibnamefont {Guerrieri}}, \bibinfo {author} {\bibfnamefont {Alexandre}\
  \bibnamefont {Homrich}}, \ and\ \bibinfo {author} {\bibfnamefont {Pedro}\
  \bibnamefont {Vieira}},\ }\bibfield  {title} {\enquote {\bibinfo {title}
  {{Dual S-matrix bootstrap. Part I. 2D theory}},}\ }\href {\doibase
  10.1007/JHEP11(2020)084} {\bibfield  {journal} {\bibinfo  {journal} {JHEP}\
  }\textbf {\bibinfo {volume} {11}},\ \bibinfo {pages} {084} (\bibinfo {year}
  {2020}{\natexlab{b}})},\ \Eprint {http://arxiv.org/abs/2008.02770}
  {arXiv:2008.02770 [hep-th]} \BibitemShut {NoStop}%
\bibitem [{\citenamefont {He}\ and\ \citenamefont {Kruczenski}()}]{talkMartin}%
  \BibitemOpen
  \bibfield  {author} {\bibinfo {author} {\bibfnamefont {Y.}~\bibnamefont
  {He}}\ and\ \bibinfo {author} {\bibfnamefont {M.}~\bibnamefont
  {Kruczenski}},\ }\href@noop {} {\bibinfo  {journal} {{to appear. See also
  talk by M.~Kruczenski at the Bootstrap 2020 annual conference in June 2020 in
  Boston (via Zoom)}}\ }\BibitemShut {NoStop}%
\bibitem [{\citenamefont {Blas}\ \emph {et~al.}(2020)\citenamefont {Blas},
  \citenamefont {Martin~Camalich},\ and\ \citenamefont {Oller}}]{gravipaper}%
  \BibitemOpen
\bibfield  {journal} {  }\bibfield  {author} {\bibinfo {author} {\bibfnamefont
  {Diego}\ \bibnamefont {Blas}}, \bibinfo {author} {\bibfnamefont {Jorge}\
  \bibnamefont {Martin~Camalich}}, \ and\ \bibinfo {author} {\bibfnamefont
  {Jose~Antonio}\ \bibnamefont {Oller}},\ }\bibfield  {title} {\enquote
  {\bibinfo {title} {{Unitarization of infinite-range forces: graviton-graviton
  scattering}},}\ }\href@noop {} {\  (\bibinfo {year} {2020})},\ \Eprint
  {http://arxiv.org/abs/2010.12459} {arXiv:2010.12459 [hep-th]} \BibitemShut
  {NoStop}%
\bibitem [{\citenamefont {Aprile}\ \emph {et~al.}(2018)\citenamefont {Aprile},
  \citenamefont {Drummond}, \citenamefont {Heslop},\ and\ \citenamefont
  {Paul}}]{Aprile}%
  \BibitemOpen
  \bibfield  {author} {\bibinfo {author} {\bibfnamefont {F.}~\bibnamefont
  {Aprile}}, \bibinfo {author} {\bibfnamefont {J.~M.}\ \bibnamefont
  {Drummond}}, \bibinfo {author} {\bibfnamefont {P.}~\bibnamefont {Heslop}}, \
  and\ \bibinfo {author} {\bibfnamefont {H.}~\bibnamefont {Paul}},\ }\bibfield
  {title} {\enquote {\bibinfo {title} {{Quantum Gravity from Conformal Field
  Theory}},}\ }\href {\doibase 10.1007/JHEP01(2018)035} {\bibfield  {journal}
  {\bibinfo  {journal} {JHEP}\ }\textbf {\bibinfo {volume} {01}},\ \bibinfo
  {pages} {035} (\bibinfo {year} {2018})},\ \Eprint
  {http://arxiv.org/abs/1706.02822} {arXiv:1706.02822 [hep-th]} \BibitemShut
  {NoStop}%
\bibitem [{\citenamefont {Alday}\ and\ \citenamefont
  {Caron-Huot}(2018)}]{Alday:2017vkk}%
  \BibitemOpen
  \bibfield  {author} {\bibinfo {author} {\bibfnamefont {Luis~F.}\ \bibnamefont
  {Alday}}\ and\ \bibinfo {author} {\bibfnamefont {Simon}\ \bibnamefont
  {Caron-Huot}},\ }\bibfield  {title} {\enquote {\bibinfo {title}
  {{Gravitational S-matrix from CFT dispersion relations}},}\ }\href {\doibase
  10.1007/JHEP12(2018)017} {\bibfield  {journal} {\bibinfo  {journal} {JHEP}\
  }\textbf {\bibinfo {volume} {12}},\ \bibinfo {pages} {017} (\bibinfo {year}
  {2018})},\ \Eprint {http://arxiv.org/abs/1711.02031} {arXiv:1711.02031
  [hep-th]} \BibitemShut {NoStop}%
\bibitem [{\citenamefont {Aharony}\ \emph {et~al.}(2017)\citenamefont
  {Aharony}, \citenamefont {Alday}, \citenamefont {Bissi},\ and\ \citenamefont
  {Perlmutter}}]{Aharony:2016dwx}%
  \BibitemOpen
  \bibfield  {author} {\bibinfo {author} {\bibfnamefont {Ofer}\ \bibnamefont
  {Aharony}}, \bibinfo {author} {\bibfnamefont {Luis~F.}\ \bibnamefont
  {Alday}}, \bibinfo {author} {\bibfnamefont {Agnese}\ \bibnamefont {Bissi}}, \
  and\ \bibinfo {author} {\bibfnamefont {Eric}\ \bibnamefont {Perlmutter}},\
  }\bibfield  {title} {\enquote {\bibinfo {title} {{Loops in AdS from Conformal
  Field Theory}},}\ }\href {\doibase 10.1007/JHEP07(2017)036} {\bibfield
  {journal} {\bibinfo  {journal} {JHEP}\ }\textbf {\bibinfo {volume} {07}},\
  \bibinfo {pages} {036} (\bibinfo {year} {2017})},\ \Eprint
  {http://arxiv.org/abs/1612.03891} {arXiv:1612.03891 [hep-th]} \BibitemShut
  {NoStop}%
\bibitem [{\citenamefont {Okuda}\ and\ \citenamefont
  {Penedones}(2011)}]{joaoYM}%
  \BibitemOpen
  \bibfield  {author} {\bibinfo {author} {\bibfnamefont {Takuya}\ \bibnamefont
  {Okuda}}\ and\ \bibinfo {author} {\bibfnamefont {Joao}\ \bibnamefont
  {Penedones}},\ }\bibfield  {title} {\enquote {\bibinfo {title} {{String
  scattering in flat space and a scaling limit of Yang-Mills correlators}},}\
  }\href {\doibase 10.1103/PhysRevD.83.086001} {\bibfield  {journal} {\bibinfo
  {journal} {Phys. Rev. D}\ }\textbf {\bibinfo {volume} {83}},\ \bibinfo
  {pages} {086001} (\bibinfo {year} {2011})},\ \Eprint
  {http://arxiv.org/abs/1002.2641} {arXiv:1002.2641 [hep-th]} \BibitemShut
  {NoStop}%
\bibitem [{\citenamefont {Chester}\ and\ \citenamefont
  {Pufu}(2021)}]{Chester:2020dja}%
  \BibitemOpen
  \bibfield  {author} {\bibinfo {author} {\bibfnamefont {Shai~M.}\ \bibnamefont
  {Chester}}\ and\ \bibinfo {author} {\bibfnamefont {Silviu~S.}\ \bibnamefont
  {Pufu}},\ }\bibfield  {title} {\enquote {\bibinfo {title} {{Far beyond the
  planar limit in strongly-coupled $ \mathcal{N} $ = 4 SYM}},}\ }\href
  {\doibase 10.1007/JHEP01(2021)103} {\bibfield  {journal} {\bibinfo  {journal}
  {JHEP}\ }\textbf {\bibinfo {volume} {01}},\ \bibinfo {pages} {103} (\bibinfo
  {year} {2021})},\ \Eprint {http://arxiv.org/abs/2003.08412} {arXiv:2003.08412
  [hep-th]} \BibitemShut {NoStop}%
\bibitem [{\citenamefont {Chester}\ \emph
  {et~al.}(2020{\natexlab{b}})\citenamefont {Chester}, \citenamefont {Green},
  \citenamefont {Pufu}, \citenamefont {Wang},\ and\ \citenamefont
  {Wen}}]{Chester:2020vyz}%
  \BibitemOpen
  \bibfield  {author} {\bibinfo {author} {\bibfnamefont {Shai~M.}\ \bibnamefont
  {Chester}}, \bibinfo {author} {\bibfnamefont {Michael~B.}\ \bibnamefont
  {Green}}, \bibinfo {author} {\bibfnamefont {Silviu~S.}\ \bibnamefont {Pufu}},
  \bibinfo {author} {\bibfnamefont {Yifan}\ \bibnamefont {Wang}}, \ and\
  \bibinfo {author} {\bibfnamefont {Congkao}\ \bibnamefont {Wen}},\ }\bibfield
  {title} {\enquote {\bibinfo {title} {{New Modular Invariants in
  $\mathcal{N}=4$ Super-Yang-Mills Theory}},}\ }\href@noop {} {\  (\bibinfo
  {year} {2020}{\natexlab{b}})},\ \Eprint {http://arxiv.org/abs/2008.02713}
  {arXiv:2008.02713 [hep-th]} \BibitemShut {NoStop}%
\bibitem [{\citenamefont {Beem}\ \emph {et~al.}(2013)\citenamefont {Beem},
  \citenamefont {Rastelli},\ and\ \citenamefont {van Rees}}]{Beem:2013qxa}%
  \BibitemOpen
  \bibfield  {author} {\bibinfo {author} {\bibfnamefont {Christopher}\
  \bibnamefont {Beem}}, \bibinfo {author} {\bibfnamefont {Leonardo}\
  \bibnamefont {Rastelli}}, \ and\ \bibinfo {author} {\bibfnamefont {Balt~C.}\
  \bibnamefont {van Rees}},\ }\bibfield  {title} {\enquote {\bibinfo {title}
  {{The $\mathcal N=4$ Superconformal Bootstrap}},}\ }\href {\doibase
  10.1103/PhysRevLett.111.071601} {\bibfield  {journal} {\bibinfo  {journal}
  {Phys. Rev. Lett.}\ }\textbf {\bibinfo {volume} {111}},\ \bibinfo {pages}
  {071601} (\bibinfo {year} {2013})},\ \Eprint {http://arxiv.org/abs/1304.1803}
  {arXiv:1304.1803 [hep-th]} \BibitemShut {NoStop}%
\bibitem [{\citenamefont {Alday}\ and\ \citenamefont
  {Bissi}(2014{\natexlab{a}})}]{Alday:2013opa}%
  \BibitemOpen
  \bibfield  {author} {\bibinfo {author} {\bibfnamefont {Luis~F.}\ \bibnamefont
  {Alday}}\ and\ \bibinfo {author} {\bibfnamefont {Agnese}\ \bibnamefont
  {Bissi}},\ }\bibfield  {title} {\enquote {\bibinfo {title} {{The
  superconformal bootstrap for structure constants}},}\ }\href {\doibase
  10.1007/JHEP09(2014)144} {\bibfield  {journal} {\bibinfo  {journal} {JHEP}\
  }\textbf {\bibinfo {volume} {09}},\ \bibinfo {pages} {144} (\bibinfo {year}
  {2014}{\natexlab{a}})},\ \Eprint {http://arxiv.org/abs/1310.3757}
  {arXiv:1310.3757 [hep-th]} \BibitemShut {NoStop}%
\bibitem [{\citenamefont {Chester}\ \emph {et~al.}(2014)\citenamefont
  {Chester}, \citenamefont {Lee}, \citenamefont {Pufu},\ and\ \citenamefont
  {Yacoby}}]{Chester:2014fya}%
  \BibitemOpen
  \bibfield  {author} {\bibinfo {author} {\bibfnamefont {Shai~M.}\ \bibnamefont
  {Chester}}, \bibinfo {author} {\bibfnamefont {Jaehoon}\ \bibnamefont {Lee}},
  \bibinfo {author} {\bibfnamefont {Silviu~S.}\ \bibnamefont {Pufu}}, \ and\
  \bibinfo {author} {\bibfnamefont {Ran}\ \bibnamefont {Yacoby}},\ }\bibfield
  {title} {\enquote {\bibinfo {title} {{The $ \mathcal{N}=8 $ superconformal
  bootstrap in three dimensions}},}\ }\href {\doibase 10.1007/JHEP09(2014)143}
  {\bibfield  {journal} {\bibinfo  {journal} {JHEP}\ }\textbf {\bibinfo
  {volume} {09}},\ \bibinfo {pages} {143} (\bibinfo {year} {2014})},\ \Eprint
  {http://arxiv.org/abs/1406.4814} {arXiv:1406.4814 [hep-th]} \BibitemShut
  {NoStop}%
\bibitem [{\citenamefont {Beem}\ \emph {et~al.}(2016)\citenamefont {Beem},
  \citenamefont {Lemos}, \citenamefont {Rastelli},\ and\ \citenamefont {van
  Rees}}]{Beem:2015aoa}%
  \BibitemOpen
  \bibfield  {author} {\bibinfo {author} {\bibfnamefont {Christopher}\
  \bibnamefont {Beem}}, \bibinfo {author} {\bibfnamefont {Madalena}\
  \bibnamefont {Lemos}}, \bibinfo {author} {\bibfnamefont {Leonardo}\
  \bibnamefont {Rastelli}}, \ and\ \bibinfo {author} {\bibfnamefont {Balt~C.}\
  \bibnamefont {van Rees}},\ }\bibfield  {title} {\enquote {\bibinfo {title}
  {{The (2, 0) superconformal bootstrap}},}\ }\href {\doibase
  10.1103/PhysRevD.93.025016} {\bibfield  {journal} {\bibinfo  {journal} {Phys.
  Rev. D}\ }\textbf {\bibinfo {volume} {93}},\ \bibinfo {pages} {025016}
  (\bibinfo {year} {2016})},\ \Eprint {http://arxiv.org/abs/1507.05637}
  {arXiv:1507.05637 [hep-th]} \BibitemShut {NoStop}%
\bibitem [{\citenamefont {Beem}\ \emph {et~al.}(2017)\citenamefont {Beem},
  \citenamefont {Rastelli},\ and\ \citenamefont {van Rees}}]{Beem:2016wfs}%
  \BibitemOpen
  \bibfield  {author} {\bibinfo {author} {\bibfnamefont {Christopher}\
  \bibnamefont {Beem}}, \bibinfo {author} {\bibfnamefont {Leonardo}\
  \bibnamefont {Rastelli}}, \ and\ \bibinfo {author} {\bibfnamefont {Balt~C.}\
  \bibnamefont {van Rees}},\ }\bibfield  {title} {\enquote {\bibinfo {title}
  {{More ${\mathcal N}=4$ superconformal bootstrap}},}\ }\href {\doibase
  10.1103/PhysRevD.96.046014} {\bibfield  {journal} {\bibinfo  {journal} {Phys.
  Rev. D}\ }\textbf {\bibinfo {volume} {96}},\ \bibinfo {pages} {046014}
  (\bibinfo {year} {2017})},\ \Eprint {http://arxiv.org/abs/1612.02363}
  {arXiv:1612.02363 [hep-th]} \BibitemShut {NoStop}%
\bibitem [{\citenamefont {Agmon}\ \emph {et~al.}(2018)\citenamefont {Agmon},
  \citenamefont {Chester},\ and\ \citenamefont {Pufu}}]{Agmon:2017xes}%
  \BibitemOpen
  \bibfield  {author} {\bibinfo {author} {\bibfnamefont {Nathan~B.}\
  \bibnamefont {Agmon}}, \bibinfo {author} {\bibfnamefont {Shai~M.}\
  \bibnamefont {Chester}}, \ and\ \bibinfo {author} {\bibfnamefont {Silviu~S.}\
  \bibnamefont {Pufu}},\ }\bibfield  {title} {\enquote {\bibinfo {title}
  {{Solving M-theory with the Conformal Bootstrap}},}\ }\href {\doibase
  10.1007/JHEP06(2018)159} {\bibfield  {journal} {\bibinfo  {journal} {JHEP}\
  }\textbf {\bibinfo {volume} {06}},\ \bibinfo {pages} {159} (\bibinfo {year}
  {2018})},\ \Eprint {http://arxiv.org/abs/1711.07343} {arXiv:1711.07343
  [hep-th]} \BibitemShut {NoStop}%
\bibitem [{\citenamefont {Arkani-Hamed}\ \emph {et~al.}(2017)\citenamefont
  {Arkani-Hamed}, \citenamefont {Huang},\ and\ \citenamefont
  {Huang}}]{Arkani-Hamed:2017jhn}%
  \BibitemOpen
  \bibfield  {author} {\bibinfo {author} {\bibfnamefont {Nima}\ \bibnamefont
  {Arkani-Hamed}}, \bibinfo {author} {\bibfnamefont {Tzu-Chen}\ \bibnamefont
  {Huang}}, \ and\ \bibinfo {author} {\bibfnamefont {Yu-tin}\ \bibnamefont
  {Huang}},\ }\bibfield  {title} {\enquote {\bibinfo {title} {{Scattering
  Amplitudes For All Masses and Spins}},}\ }\href@noop {} {\  (\bibinfo {year}
  {2017})},\ \Eprint {http://arxiv.org/abs/1709.04891} {arXiv:1709.04891
  [hep-th]} \BibitemShut {NoStop}%
\bibitem [{\citenamefont {Chowdhury}\ \emph {et~al.}(2020)\citenamefont
  {Chowdhury}, \citenamefont {Gadde}, \citenamefont {Gopalka}, \citenamefont
  {Halder}, \citenamefont {Janagal},\ and\ \citenamefont
  {Minwalla}}]{Chowdhury:2019kaq}%
  \BibitemOpen
  \bibfield  {author} {\bibinfo {author} {\bibfnamefont {Subham~Dutta}\
  \bibnamefont {Chowdhury}}, \bibinfo {author} {\bibfnamefont {Abhijit}\
  \bibnamefont {Gadde}}, \bibinfo {author} {\bibfnamefont {Tushar}\
  \bibnamefont {Gopalka}}, \bibinfo {author} {\bibfnamefont {Indranil}\
  \bibnamefont {Halder}}, \bibinfo {author} {\bibfnamefont {Lavneet}\
  \bibnamefont {Janagal}}, \ and\ \bibinfo {author} {\bibfnamefont {Shiraz}\
  \bibnamefont {Minwalla}},\ }\bibfield  {title} {\enquote {\bibinfo {title}
  {{Classifying and constraining local four photon and four graviton
  S-matrices}},}\ }\href {\doibase 10.1007/JHEP02(2020)114} {\bibfield
  {journal} {\bibinfo  {journal} {JHEP}\ }\textbf {\bibinfo {volume} {02}},\
  \bibinfo {pages} {114} (\bibinfo {year} {2020})},\ \Eprint
  {http://arxiv.org/abs/1910.14392} {arXiv:1910.14392 [hep-th]} \BibitemShut
  {NoStop}%
\bibitem [{\citenamefont {Bern}\ \emph {et~al.}(1998)\citenamefont {Bern},
  \citenamefont {Dixon}, \citenamefont {Dunbar}, \citenamefont {Perelstein},\
  and\ \citenamefont {Rozowsky}}]{Bern:1998ug}%
  \BibitemOpen
  \bibfield  {author} {\bibinfo {author} {\bibfnamefont {Z.}~\bibnamefont
  {Bern}}, \bibinfo {author} {\bibfnamefont {Lance~J.}\ \bibnamefont {Dixon}},
  \bibinfo {author} {\bibfnamefont {D.~C.}\ \bibnamefont {Dunbar}}, \bibinfo
  {author} {\bibfnamefont {M.}~\bibnamefont {Perelstein}}, \ and\ \bibinfo
  {author} {\bibfnamefont {J.~S.}\ \bibnamefont {Rozowsky}},\ }\bibfield
  {title} {\enquote {\bibinfo {title} {{On the relationship between Yang-Mills
  theory and gravity and its implication for ultraviolet divergences}},}\
  }\href {\doibase 10.1016/S0550-3213(98)00420-9} {\bibfield  {journal}
  {\bibinfo  {journal} {Nucl. Phys. B}\ }\textbf {\bibinfo {volume} {530}},\
  \bibinfo {pages} {401--456} (\bibinfo {year} {1998})},\ \Eprint
  {http://arxiv.org/abs/hep-th/9802162} {arXiv:hep-th/9802162} \BibitemShut
  {NoStop}%
\bibitem [{\citenamefont {Green}\ \emph {et~al.}(2008)\citenamefont {Green},
  \citenamefont {Russo},\ and\ \citenamefont {Vanhove}}]{Green:2008uj}%
  \BibitemOpen
  \bibfield  {author} {\bibinfo {author} {\bibfnamefont {Michael~B.}\
  \bibnamefont {Green}}, \bibinfo {author} {\bibfnamefont {Jorge~G.}\
  \bibnamefont {Russo}}, \ and\ \bibinfo {author} {\bibfnamefont {Pierre}\
  \bibnamefont {Vanhove}},\ }\bibfield  {title} {\enquote {\bibinfo {title}
  {{Low energy expansion of the four-particle genus-one amplitude in type II
  superstring theory}},}\ }\href {\doibase 10.1088/1126-6708/2008/02/020}
  {\bibfield  {journal} {\bibinfo  {journal} {JHEP}\ }\textbf {\bibinfo
  {volume} {02}},\ \bibinfo {pages} {020} (\bibinfo {year} {2008})},\ \Eprint
  {http://arxiv.org/abs/0801.0322} {arXiv:0801.0322 [hep-th]} \BibitemShut
  {NoStop}%
\bibitem [{\citenamefont {Caron-Huot}\ and\ \citenamefont
  {Trinh}(2019)}]{Caron-Huot:2018kta}%
  \BibitemOpen
  \bibfield  {author} {\bibinfo {author} {\bibfnamefont {Simon}\ \bibnamefont
  {Caron-Huot}}\ and\ \bibinfo {author} {\bibfnamefont {Anh-Khoi}\ \bibnamefont
  {Trinh}},\ }\bibfield  {title} {\enquote {\bibinfo {title} {{All tree-level
  correlators in AdS$_{5}\times{}S_{5}$ supergravity: hidden ten-dimensional
  conformal symmetry}},}\ }\href {\doibase 10.1007/JHEP01(2019)196} {\bibfield
  {journal} {\bibinfo  {journal} {JHEP}\ }\textbf {\bibinfo {volume} {01}},\
  \bibinfo {pages} {196} (\bibinfo {year} {2019})},\ \Eprint
  {http://arxiv.org/abs/1809.09173} {arXiv:1809.09173 [hep-th]} \BibitemShut
  {NoStop}%
\bibitem [{\citenamefont {Martin}(1963)}]{Martin:1962rt}%
  \BibitemOpen
  \bibfield  {author} {\bibinfo {author} {\bibfnamefont {A.}~\bibnamefont
  {Martin}},\ }\bibfield  {title} {\enquote {\bibinfo {title} {{Unitarity and
  high-energy behavior of scattering amplitudes}},}\ }\href {\doibase
  10.1103/PhysRev.129.1432} {\bibfield  {journal} {\bibinfo  {journal} {Phys.
  Rev.}\ }\textbf {\bibinfo {volume} {129}},\ \bibinfo {pages} {1432--1436}
  (\bibinfo {year} {1963})}\BibitemShut {NoStop}%
\bibitem [{\citenamefont {Camanho}\ \emph {et~al.}(2016)\citenamefont
  {Camanho}, \citenamefont {Edelstein}, \citenamefont {Maldacena},\ and\
  \citenamefont {Zhiboedov}}]{Camanho:2014apa}%
  \BibitemOpen
  \bibfield  {author} {\bibinfo {author} {\bibfnamefont {Xian~O.}\ \bibnamefont
  {Camanho}}, \bibinfo {author} {\bibfnamefont {Jose~D.}\ \bibnamefont
  {Edelstein}}, \bibinfo {author} {\bibfnamefont {Juan}\ \bibnamefont
  {Maldacena}}, \ and\ \bibinfo {author} {\bibfnamefont {Alexander}\
  \bibnamefont {Zhiboedov}},\ }\bibfield  {title} {\enquote {\bibinfo {title}
  {{Causality Constraints on Corrections to the Graviton Three-Point
  Coupling}},}\ }\href {\doibase 10.1007/JHEP02(2016)020} {\bibfield  {journal}
  {\bibinfo  {journal} {JHEP}\ }\textbf {\bibinfo {volume} {02}},\ \bibinfo
  {pages} {020} (\bibinfo {year} {2016})},\ \Eprint
  {http://arxiv.org/abs/1407.5597} {arXiv:1407.5597 [hep-th]} \BibitemShut
  {NoStop}%
\bibitem [{\citenamefont {Alday}\ and\ \citenamefont
  {Bissi}(2014{\natexlab{b}})}]{AldayBissi}%
  \BibitemOpen
  \bibfield  {author} {\bibinfo {author} {\bibfnamefont {Luis~F.}\ \bibnamefont
  {Alday}}\ and\ \bibinfo {author} {\bibfnamefont {Agnese}\ \bibnamefont
  {Bissi}},\ }\bibfield  {title} {\enquote {\bibinfo {title} {{Modular
  interpolating functions for N=4 SYM}},}\ }\href {\doibase
  10.1007/JHEP07(2014)007} {\bibfield  {journal} {\bibinfo  {journal} {JHEP}\
  }\textbf {\bibinfo {volume} {07}},\ \bibinfo {pages} {007} (\bibinfo {year}
  {2014}{\natexlab{b}})},\ \Eprint {http://arxiv.org/abs/1311.3215}
  {arXiv:1311.3215 [hep-th]} \BibitemShut {NoStop}%
\bibitem [{\citenamefont {Truong}(1988)}]{Truong:1988zp}%
  \BibitemOpen
  \bibfield  {author} {\bibinfo {author} {\bibfnamefont {Tran~N.}\ \bibnamefont
  {Truong}},\ }\bibfield  {title} {\enquote {\bibinfo {title} {{Chiral
  Perturbation Theory and Final State Theorem}},}\ }\href {\doibase
  10.1103/PhysRevLett.61.2526} {\bibfield  {journal} {\bibinfo  {journal}
  {Phys. Rev. Lett.}\ }\textbf {\bibinfo {volume} {61}},\ \bibinfo {pages}
  {2526} (\bibinfo {year} {1988})}\BibitemShut {NoStop}%
\bibitem [{\citenamefont {Dobado}\ and\ \citenamefont
  {Pelaez}(1997)}]{Dobado:1996ps}%
  \BibitemOpen
  \bibfield  {author} {\bibinfo {author} {\bibfnamefont {A.}~\bibnamefont
  {Dobado}}\ and\ \bibinfo {author} {\bibfnamefont {J.~R.}\ \bibnamefont
  {Pelaez}},\ }\bibfield  {title} {\enquote {\bibinfo {title} {{The Inverse
  amplitude method in chiral perturbation theory}},}\ }\href {\doibase
  10.1103/PhysRevD.56.3057} {\bibfield  {journal} {\bibinfo  {journal} {Phys.
  Rev. D}\ }\textbf {\bibinfo {volume} {56}},\ \bibinfo {pages} {3057--3073}
  (\bibinfo {year} {1997})},\ \Eprint {http://arxiv.org/abs/hep-ph/9604416}
  {arXiv:hep-ph/9604416} \BibitemShut {NoStop}%
\end{thebibliography}%

\end{document}